\documentclass{aa}  

\usepackage{graphicx}
\usepackage{pifont}
\usepackage[colorlinks=true,citecolor=blue]{hyperref}

\usepackage{txfonts}
\begin{document} 
\defcitealias{heinze2023}{H23}
\defcitealias{despali22a}{Paper I}

   \title{Detecting low-mass haloes with strong gravitational lensing II: \\ constraints on the density profiles of two detected subhaloes}
   \titlerunning{Density profiles of subhalo detections}


   \author{Giulia Despali\inst{1,2,3}\fnmsep\thanks{giulia.despali@unibo.it} and Felix M. Heinze\inst{4}\fnmsep\thanks{felix.heinze@uni-jena.de}
          ,
        \\  Christopher D. Fassnacht\inst{5}, Simona Vegetti\inst{6}, Cristiana Spingola\inst{7}, Ralf Klessen\inst{8,9}
          }
    \authorrunning{Despali, Heinze et al.}

        \institute{
        Dipartimento di Fisica e Astronomia "Augusto Righi", Alma Mater Studiorum Università di Bologna, via Gobetti 93/2, I-40129 Bologna, Italy
         \and
         INAF-Osservatorio di Astrofisica e Scienza dello Spazio di Bologna, Via Piero Gobetti 93/3, I-40129 Bologna, Italy
         \and
        INFN-Sezione di Bologna, Viale Berti Pichat 6/2, I-40127 Bologna, Italy
        \and 
        Friedrich-Schiller-Universität Jena, Theoretisch-Physikalisches Institut, Fröbelstieg 1, D-07743 Jena, Germany
        \and
        Department of Physics and Astronomy, University of California Davis, 1 Shields Avenue, Davis, CA 95616, USA
        \and
        Max Planck Institute for Astrophysics, Karl-Schwarzschild Str. 1, D-85748 Garching bei M\"unchen, Germany
        \and 
        INAF $-$ Istituto di Radioastronomia, via Gobetti 101, I-40129 Bologna, Italy
        \and
        Universität Heidelberg, Zentrum für Astronomie, Institut für Theoretische Astrophysik, Albert-Ueberle-Straße 2, D-69120 Heidelberg, Germany
        \and
        Universität Heidelberg, Interdisziplinäres Zentrum für Wissenschaftliches Rechnen, Im Neuenheimer Feld 205, D-69120 Heidelberg, Germany
             }

   \date{Submitted 07/2024}

 
  \abstract
  {}
   {Strong gravitational lensing can detect the presence of low-mass haloes and subhaloes through their effect on the surface brightness of lensed arcs. We carry out an extended analysis of the density profiles and mass distributions of two detected subhaloes, intending to determine if their properties are consistent with the predictions of the cold dark matter (CDM) model.}
   {We analyse two gravitational lensing systems in which the presence of two low-mass subhaloes has been previously reported: SDSSJ0946+1006 and JVASB1938+66. We model these detections assuming four different models for their density profiles and compare our results with predictions from the IllustrisTNG50-1 simulation.}
   { We find that the detected subhaloes are well-modelled by steep inner density slopes, close to or steeper than isothermal. The NFW profile thus needs extremely high concentrations to reproduce the observed properties, which are outliers of the CDM predictions. We also find a characteristic radius within which the best-fitting density profiles predict the same enclosed mass. We conclude that the lens modelling can constrain this quantity more robustly than the inner slope. We find that the diversity of subhalo profiles in TNG50, consistent with tidally stripping and baryonic effects, is able to match the observed steep inner slopes, somewhat alleviating the tension reported by previous works even if the detections are not well fit by the typical subhalo. However, while we find simulated analogues of the detection in B1938+666, the stellar content required by simulations to explain the central density of the detection in J0946+1006 is in tension with the upper limit in luminosity estimated from the observations. New detections will increase our statistical sample and help us reveal more about the density profiles of these objects and the dark matter content of the Universe.
   }
   {}


   \maketitle
%

\section{Introduction}

Strong gravitational lensing is one of the most promising methods for studying the nature of dark matter. It allows one to detect low-mass dark subhaloes within the haloes of lens galaxies and isolated dark haloes along their line of sight, providing a quantitative test of the Cold Dark Matter (CDM) paradigm in a halo mass regime and for distances that are not easily accessible to other techniques. It also offers a robust method to distinguish between CDM and alternative models in which the abundance of low-mass haloes is suppressed, for example, Warm Dark Matter \citep[WDM, e.g.][]{schneider12,lovell12,lovell14,despali20} and Self-Interacting Dark Matter \citep[SIDM, e.g.][]{vogel12,vogel16,despali19}, or models that create small-scale features such as the granules in Fuzzy Dark Matter \citep[FDM, e.g.][]{robles17,powell23}. Given that the distortion of lensed images is due to gravity only, we can directly measure the total mass of the object acting as a lens independently of its dark or baryonic nature, both in the case of the main lens galaxy (typically an Early-Type galaxy) and additional smaller low-mass haloes. However, see \citet{vegetti23} for a discussion on systematic errors induced by baryonic effects. 

The search for dark matter haloes and subhaloes through their effect on magnified arcs and Einstein rings has led to three detections of low-mass dark matter (sub)haloes. A first object of mass $M_{\rm sub}=(3.51\pm0.15)\times10^{9}$ M$_{\odot}$ in the Hubble Space Telescope \citep[HST,][]{vegetti10} observations of the system J0946+1006 and a second perturber of mass $M_{\rm sub}=(1.9\pm0.1)\times10^{8}$ M$_{\odot}$ using Keck adaptive optics imaging \citep{vegetti12} of B1938+666. Moreover, one detection $M_{\rm sub}=10^{8.96\pm0.12}$ M$_{\odot}$ was reported by \citet{hezaveh16} in ALMA observations of the system SDP.81. In all cases, the subhalo mass corresponds to a Pseudo-Jaffe profile \citep{jaffe83,munoz01}, i.e. a steeply-truncated isothermal profile. The first two results, combined with the information from non-detections, have been used to set constraints on the particle mass of WDM models \citep{vegetti18,ritondale19b,enzi21}. In practice, obtaining reliable and precise forecasts for the number and properties of gravitational lens systems required for more stringent constraints is challenging. As demonstrated in the first paper of this series \citep[][hereafter Paper I]{despali22a}, several factors influence the data sensitivity: the angular resolution of the instrument, the redshift of the lens and source galaxies, the details of the surface brightness distribution of the source, the size of the lensed images, the fidelity of the observed images (e.g. SNR) and the threshold used to define a detection. The results of \citetalias{despali22a} were based on the assumption that dark haloes and subhaloes are well described by the classic NFW profile \citep{navarro96} predicted by dark matter simulations, which was not the profile adopted by the observational analyses that led to the detections. Recent works by \citet{minor2021}, \citet{sengul22} and \citet{ballard24} reanalysed the HST observations of the first two systems using an NFW profile to describe the perturbers, following the predictions of numerical simulations. The inferred NFW concentrations required to explain the observed deflections are substantially above the halo concentration-mass relation measured in simulations \citep{duffy08,ludlow16,moline17}. Moreover, \citet{sengul22} claimed that the detection in B1938+666 is best explained by an isolated dark halo along the line of sight, located behind the main lens at $z=1.4$, rather than a subhalo. \citet{minor2021} compared their results on J0946+1006 to the subhalo profiles from the IllustrisTNG-100-1 hydrodynamical simulation \citep{pillepich18}, finding that the inferred subhalo properties are outliers of the simulated distribution and thus possibly of the CDM model. 

Analyses carried out at other scales report similar trends in the properties of detected subhaloes. For example, \citet{bonaca19} modelled the Milky-Way stellar stream GD-1 to constrain the subhalo mass that could have generated the gap and spur in the observed distribution of stars. They concluded that the observed properties of the stream are consistent with a past interaction with a subhalo of mass $10^{6}-10^{8}$ M$_{\odot}$ and an NFW scale radius $r_\mathrm{s}=20$ pc. The subhalo size is smaller than predicted by numerical simulations that resolve these scales and in which scale radii are of the order of 50 pc \citep{moline17}, thus implying a high concentration. At larger scales, satellite galaxies in clusters are found to be more efficient lenses than their counterparts in numerical simulations \citep{meneghetti20,meneghetti23,ragagnin22,ragagnin23}, pointing in the same direction. Power-law profiles with slopes steeper than isothermal have also been found for main lensing galaxies, especially those belonging to groups or clusters of galaxies \citep{rusin02,dobke07,spingola18}. Other than individual cases, within the SLACS sample, it has been found that those lenses with a perturbing companion galaxy are much better modelled with a super-isothermal mass density profile, suggesting that the steep slope is likely caused by interactions \citep{auger08}. It is thus not surprising that the presence of baryons and tidal effects (or both) may lead to inner profile slopes steeper than the classic NFW model.

These results motivate this work and make us wonder if the current detections are effectively in tension with CDM and what role uncertainties play in predictions from simulations. One mechanism that has been proposed as a possible solution to this tension is the steepening of profiles due to gravothermal core-collapse in SIDM models. Velocity-dependent cross-sections can favour interactions at small scales and, while initially creating a core, lead to an instability which re-attracts matter to the centre to form a cuspy structure \citep{tulin18,koda11,kaplinghat19,adhikari22,yang23}. \citet{nadler21} investigated this possibility and found a better match between the steep subhalo slope measured by \citet{minor2021} and their simulation. However, it is also possible that we only detect structures belonging to a specific sub-population of particularly compact objects. For example, they could be heavily stripped subhaloes, where the density profile has been truncated by the interaction with the host but retains a central large density. Previous works, such as \cite{dicintio11, dicintio13}, already pointed out that the NFW profile might not be a suitable model for describing the density profiles of subhaloes, especially in the presence of baryons. Even dark-matter-only simulations have shown that subhaloes exhibit much higher central densities and are generally more concentrated than isolated field haloes of comparable mass \citep{ghigna00, bullock01a, kazantzidis04b, diemand07b}. Recently, \citet[][hereafter H23]{heinze2023} have studied the total density profiles of subhaloes in the TNG50-1 hydrodynamical cosmological simulation \citep{pillepich18} and proposed a new fitting function with much steeper inner and outer log-slopes compared to those of the standard NFW profile. 

In this work, we take an agnostic approach to observational data and simulations. We re-model the two perturbers detected with gravitational imaging in J0946+1006 and B1938+666, using four different models for the density profile that differ in terms of slope and mass definition. We evaluate their performance at reproducing the observations and then compare the inferred parameters to the simulated subhaloes in the TNG50-1 run. In the latter, we measure the total density profiles, considering both the dark matter and baryonic content of subhaloes: even if substructures detected with gravitational imaging are often referred to as \emph{dark}, they may still contain stars or gas that are too faint to be detected, but that influence the density profile. The paper is structured as follows. In Section \ref{sec:method}, we outline the lensing methodology and the considered density profiles and summarise the results from \citetalias{heinze2023}. In Section \ref{sec:lens_model}, we instead describe the observational data used in this work: the systems SDSSJ0946+1006 and JVASB1938+666. In Section \ref{sec:results}, we present the lens modelling results and the corresponding inference of subhalo properties, which we then compare to simulated subhaloes in Section \ref{sec:comparison}. Finally, in Section \ref{sec:conc}, we discuss our results and present our conclusions.

\begin{figure*}
\centering
\includegraphics[width=0.47\textwidth]{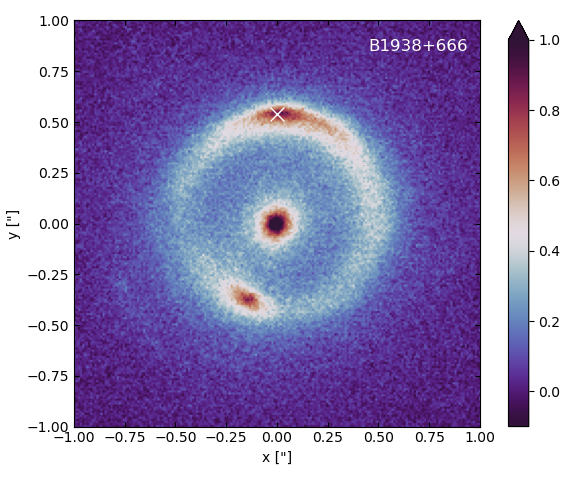}
\includegraphics[width=0.45\textwidth]{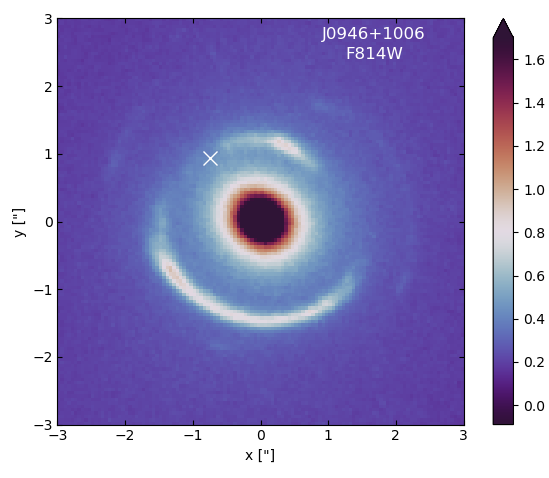}
\caption{Images of the two strong gravitational lensing systems used in this paper, where subhalo detections (marked by a white cross) have been claimed in previous works. $Left$: image of B1938+666 taken in the K$^\prime$ band by NIRC2 on Keck. In this case, $z_\mathrm{L}=0.881$ and $z_\mathrm{s}=2.059$; the pixel and FWHM (determining the resolution) scales are 10 and 70 milli-arcseconds (hereafter, \emph{mas}). $Right:$ HST F814W observations of the lens J0946+1006. For this system $z_\mathrm{L}=0.222$ and $z_\mathrm{s}=0.609$; the pixel size and resolution are 50 and 90 \emph{mas}. See Table \ref{tab:lenses} for a list of previous analyses of these systems.}
\label{fig:lensing_data}
\end{figure*}

\begin{table*}
    \centering
    \begin{tabular}{c|c|c|c|c|c|c}
    \hline
    \textbf{Lens} & \textbf{Instrument} & \textbf{Band} & \textbf{M$_{\rm sub}$ [$10^{9}$ M$_{\odot}$]} & \textbf{c$_\mathrm{sub}$} & \textbf{Profile} & \textbf{Ref.} \\
    \hline
       B1938  & Keck/NIRC2+HST & K$^\prime$+NICMOS & 0.19$\pm$0.01 & -& PJ & V12 \\
       +666 & Keck/NIRC2 & K$^\prime$ & - & - & - & L12 \\
         & HST/NICMOS & F160W & - & - & - & K98 \\
         & HST/NICMOS & F160W & 1 & $\sim$60 ($\gamma_{\rm sub}=1.96\pm0.12$) & tNFW & S22 \\
        \hline
        J0946 & HST/ACS & F814W & - & -  & -& G08 \\
         +1006  & &  & 3.51$\pm$ 0.15 & - & PJ & V10 \\
         & &  & $5.01^{+1.4}_{-1}$ , $30.2^{+9.8}_{-17.2}$ & $1560^{+1440}_{-859}$ , $70.6^{+17.8}_{-11.9}$ & tNFW & M21\\
         & &  & $8^{+87}_{-59}$ , $812^{+164}_{-583}$ & \citet{ludlow16} & NFW & N24 \\
          &+MUSE  & +F336W & $1.58^{+2.4}_{-0.33}$ & $251^{+543}_{-125}$ & tNFW & B24 \\
         & +HST/WFC3,WFPC2  & +F(160,606,438)W & - & - & - & S12 \\
         & MUSE & & - & - & - & C20 \\
    \hline

    \end{tabular}
    \caption{Summary of most relevant works where the two systems considered here have been modelled - with or without the perturber. The data are shown in Fig. \ref{fig:lensing_data}. For B1938+666, we report values from: \citet{vegetti12}, \citet{lagattuta12}, \citet{king98} and \citet{sengul22}.  For J0946+1006: \citet{gavazzi08}, \citet{vegetti10}, \citet{minor2021}, \citet{nightingale24}, \citet{ballard24}, \citet{sonnenfeld12}  and \citet{collett20}. \citet{minor2021}, \citet{sengul22} and \citet{ballard24} use a truncated NFW profile (tNFW) instead of the classic NFW model. Additionally, \citet{minor2021} modelled the mass distribution of the main lens with and without the inclusion of multipoles, the latter leading to a higher mass and concentration for the detected subhalo. \citet{nightingale24} used a set of four variations for the main lens mass model, which determine the range of values for $M_{\rm sub}$, among which we report the minimum and maximum value.}
    \label{tab:lenses}
\end{table*}

\section{Methods} \label{sec:method}

\subsection{Lensing analysis} \label{subsec:code}
In this work, we use the \textsc{pronto} lens modelling code (Vegetti et al. in prep) originally developed by  \citet{vegetti09} and further extended by \citet{rybak15}, \citet{rizzo18}, \citet{ritondale19a} and \cite{powell21}. Here, we briefly describe the method and we refer the reader to these previous works for additional details.

In the context of Bayesian statistics, the strong lensing inference problem is best expressed in terms of the following posterior distribution:
\begin{equation}
P(\mathbf{s},\boldsymbol{\eta},\lambda_\mathrm{s}|\mathbf{d})=\frac{P(\mathbf{d}|\mathbf{\eta},\mathbf{s})P(\boldsymbol{\eta})P(\mathbf{s}|\lambda_\mathrm{s})}{P(\mathbf{d})}.
\label{eq_post}    
\end{equation}
Here, $\mathbf{d}$ is the observed surface brightness distribution of the lensed images, $\mathbf{s}$ the background-source-galaxy surface brightness, $\boldsymbol{\eta}$ a vector containing the parameters describing the lens mass and light distribution, and $\lambda_\mathrm{s}$ the source regularisation level. We represent the lens mass distribution by an elliptical power-law model, with dimensionless surface mass density given by
\begin{equation}
    \kappa(x,y)=\frac{\kappa_{0}\left(2-\frac{\gamma}{2}\right)q^{\gamma-3/2}}{2\left(q^{2}x^{2}+y^{2}\right)^{(\gamma-1)/2}}\,.
    \label{eq:lens}
\end{equation}
Hence, the unknown set of parameters $\boldsymbol{\eta}$ include the mass density normalisation $\kappa_{0}$, the radial mass-density slope $\gamma$, the axis ratio $q$ (and position angle). An isothermal profile corresponds to $\gamma=2$. In addition, an external shear component of strength $\Gamma$ and position angle $\Gamma_{\theta}$ is added to the model. The Einstein radius can be then calculated as \citep{ritondale19a,enzi20}:
\begin{equation}
R_\mathrm{ein} = \left(\frac{\kappa_{0}\left(2-\frac{\gamma}{2}\right)q^{(\gamma-2)/2}}{3-\gamma} \right)^{1/(\gamma-1)}, \label{eq:rein}
\end{equation}
meaning that for an isothermal spherical lens $\kappa_{0}$ can be interpreted as the Einstein radius. We follow  \citet{vegetti09} and model the source $\mathbf{s}$ in a pixellated regularised fashion. The brightness of each pixel in the source plane, as well as the source regularisation level $\lambda_\mathrm{s}$, are thus also free parameters of the model. 

In addition to the main lens, we also model the perturber and optimise for its structural properties (e.g. mass and concentration) and position, contained in a vector $\boldsymbol{\eta_{\rm sub}}$. The analytical models adopted to describe the density profile of the subhalo are described in the next section. In order to determine the statistical significance of the subhalo detections, we calculate the marginalised Bayesian Evidence of the perturbed model with \textsc{MultiNest} \citep{feroz13} 
\begin{equation}
E_{\rm pert} = \int P(\mathbf{d}|\boldsymbol{\eta}, \boldsymbol{\eta}_{\rm{sub}},\mathbf{s})~P(\mathbf{s}|\lambda_\mathrm{s})~P(\boldsymbol{\eta})~d\lambda_\mathrm{s} d\boldsymbol{s} d\boldsymbol{\eta}d\boldsymbol{\eta}_{\rm{sub}}\,
 \label{eq_ev_pert}
\end{equation}
and compare that to a smooth model (i.e. a lens without perturbing subhaloes)
\begin{equation}
 E_{\rm smooth} = \int P(\mathbf{d}|\boldsymbol{\eta},\mathbf{s}) ~P(\mathbf{s}|\lambda_\mathrm{s})~P(\boldsymbol{\eta})~d\lambda_\mathrm{s}d\boldsymbol{s}d\boldsymbol{\eta}.
 \label{eq_ev_smooth}
\end{equation}
Both integrals are performed assuming uniform priors on the parameters $\boldsymbol{\eta}$ and a uniform prior in logarithmic space on the source regularisation $\lambda_\mathrm{s}$. We choose the size of the priors to be the same for the two models. The Bayes factor $\Delta\log E = \log  E_{\rm pert} -  \log E_{\rm smooth}$ is used to quantify the significance of the detection, as in previous works \citep{vegetti14,ritondale19b,despali22a}. These established a threshold of $\Delta\log E \geq50$ - roughly corresponding to 10-$\sigma$ detection - as a reliable way to limit the rate of false-positive detections.

 Here, we do not include multipoles in our lens mass models and focus on the relative difference between the smooth and perturbed models. We find that a standard model including an elliptical power-law profile with external shear, together with the inclusion of the subhalo, describes the mass models reasonably well, leading to very small residuals (see Fig. \ref{fig:lensing_data1} - \ref{fig:lensing_data4} in the Appendix). However, recent works have shown that the inclusion of multipoles can influence the dark matter inference \citep{oriordan24,cohen24}, thanks to the increased complexity in the mass model of the main lens that can reduce unexplained features in the residuals \citep{powell21,nightingale24,stacey24}. In this work, we consider subhaloes within the main lens as possible sources of perturbations. Another possibility is that they are caused by a dark halo located along the line of sight \citep{despali16}. It is expected that such a perturber would be well-described by an NFW profile following the concentration-mass-redshift relation. We leave a further investigation of the effect of multipoles on these two detections and the interpretation as a line-of-sight halo to future work.


\subsection{Subhalo density profiles} \label{subsec:profiles}

Here we describe four different density profiles that we use to model the density distribution of the perturbers:
\begin{itemize}
    \item a standard (thus not truncated) NFW profile following a concentration-mass determined by \citet{duffy08}. The concentration is not constant, but is determined by the subhalo mass and fixed to the mean of the $c-m$ relation for the virial mass. The NFW density profile \citep{navarro96} is defined as
\begin{equation}
\rho(r)=\dfrac{\rho_\mathrm{s}}{\frac{r}{r_\mathrm{s}}\left(1+\frac{r}{r_\mathrm{s}}\right)^{2}}, \label{eq:nfw}
\end{equation}
where $r_\mathrm{s}$ is the scale radius (i.e. the location of slope transition), and $\rho_\mathrm{s}$ is the density normalisation. The NFW profile can also be  defined   in  terms  of  the  halo   virial  mass  $M_{\rm vir}$
\citep[i.e.  the  mass   within  the  radius  that   encloses  a  virial
 overdensity $\Delta_{\rm vir}$, defined  following][]{bryan98}, 
and a concentration related to the scale radius through  $r_\mathrm{s}=r_{\rm vir}/c_{\rm sub}$.

    \item an NFW profile with free concentration, i.e. where we treat $c_{\rm sub}$ as an extra free parameter;
    \item a Pseudo-Jaffe (PJ) profile \citep{jaffe83,munoz01}. This is defined as
\begin{equation}
\rho(r)=\dfrac{\rho_{0}r_\mathrm{t}^{4}}{r^{2}(r^{2}+r_\mathrm{t}^{2})},\label{eq:pj}
\end{equation}
where $r_\mathrm{t}$ is the   truncation    radius and $\rho_0$ is the density normalisation. 

Generally, the truncation radius is assumed to approximate well the substructure tidal radius;
given the lens parameters defined in equation (\ref{eq:lens}), we calculate the mass of the host lens at the position $r$ of the subhalo as
\begin{equation}
M_\mathrm{lens}(<R) =\frac{\Sigma_\mathrm{c}\pi\kappa_0}{2}\frac{4-\gamma}{3-\gamma}q^{(\gamma-2)/2}R^{3-\gamma}.
\end{equation}
The PJ truncation radius is then \citep{jetzer98,bellazzini04}
\begin{equation}
R_\mathrm{t} = \frac{2}{3}\left(\frac{M_\mathrm{PJ}}{2M_\mathrm{lens}}\right)^{1/3} R.
\end{equation}

Note that the definitions of $r_\mathrm{t}$ and the mass of the PJ profile are different with respect to \citet{vegetti10,vegetti12} where $M_{\rm PJ}$ was defined as the mass within $r_\mathrm{t}$ and not the total mass. However, by comparing the PJ masses inferred in this work with those of the original papers, we find that the new definition does not alter the conclusions about the physical properties of the perturber.
    \item a Power-Law (PL) density profile (not truncated) where the slope is a free parameter. This is defined as in equation (\ref{eq:lens}), assuming a spherical mass distribution ($q\equiv1$) and no external shear.
\end{itemize}

We then compare our results to numerical simulations and to the findings of \citetalias{heinze2023}, which are summarised in Sect. \ref{subsec:mnfw}.

\subsection{TNG50 and the modified NFW profile (mNFW)} \label{subsec:mnfw}

The simulated subhaloes analysed in this paper are drawn from IllustrisTNG, a suite of large-volume cosmological gravo-magnetohydrodynamical simulations \citep{pillepich18}, performed using the moving-mesh code \textsc{Arepo} \citep{springel10}. They include a large number of physical processes such as primordial and metal-line cooling, heating by the extragalactic UV background, stochastic star formation, evolution of stellar populations, feedback from supernovae and AGB stars, as well as supermassive black hole formation and feedback. 
In order to be able to reliably study the small sub-galactic scales, we make use of the TNG50-1 run \citep{pillepich18}, which has a dark matter mass resolution of $3.1 \times 10^5$ M$_\odot h^{-1}$, a baryonic mass resolution of $5.7 \times 10^4$ M$_\odot h^{-1}$ and a gravitational softening length of 0.288 comoving kpc at $z=0$. Haloes and subhaloes are identified with the \textsc{Subfind} algorithm \citep{springel05a}. We also use the merger trees constructed with the \textsc{Sublink} \citep{rodriguez15_sublink} algorithm, which are part of the data released with the TNG simulations \citep{nelson19}. These are constructed at the subhalo level by identifying progenitors and descendants of each object.  Firstly, descendant candidates are identified for each subhalo as those subhaloes in the following snapshot that have common particles with the one in question. These are then ranked based on the binding energy of the shared particles, and the descendant of a subhalo is defined as the candidate with the highest score. Knowledge of all the descendants, along with the definition of the first progenitor, uniquely determines the merger trees and thus the history of each subhalo.

We rely on the results by \citetalias{heinze2023}, which modelled the $total$ density profiles (including both baryonic and dark matter) of all TNG50-1 subhaloes with masses above $1.4 \times 10^8$ M$_\odot$ and tested various fitting functions. The best-performing model is the \emph{modified NFW profile} (hereafter \emph{mNFW}), defined by the functional form:
\begin{equation}
    \rho(r) = \frac{\rho_0}{\left( \frac{r}{r_\mathrm{s}} \right)^2 \left[ 1 + \left( \frac{r}{r_\mathrm{s}} \right)^\alpha \right]^6}.
\label{eq:modified_NFW}
\end{equation}
Here, $\rho_0$ sets the density normalisation, $r_\mathrm{s}$ is the radius at which the profile transitions between the inner and outer log-slope, and $\alpha$ controls both the outer log slope and the sharpness of the slope transition, thereby determining the shape of the density profile. \citetalias{heinze2023} showed that the mean inner slope of the total density profile is close to -2 for all subhalo masses in the range $4\times10^{8}\leq M_\mathrm{sub}/\mathrm{M_{\odot}}\leq10^{12}$. This implies that the density profile of subhaloes is (on average) close to isothermal and thus steeper in both its inner and outer parts than the standard NFW model that is commonly used to describe the dark matter component. The \emph{mNFW} profile is thus close to the PJ definition, in terms of inner slope and mass distribution. As the log-slope beyond $r_\mathrm{s}$ is very steep, most of the subhalo mass is contained within $r_\mathrm{s}$, and thus, this value corresponds, in practice, to a truncation radius. We thus do not model the detections directly with this profile, but we extend their results by measuring the lensing and observable properties of the simulated subhaloes and comparing them to those of the detections in observational data (see Sect. \ref{sec:comparison}). In this way, we can also account for the significant scatter around the mean values of slope and $\rho_{0}$ reported by \citetalias{heinze2023}, highlighting the large diversity of subhalo properties that can be expected due to the baryonic effects and tidal stripping \citep{kazantzidis04b, green19, moline23}.

\subsection{A note on units} \label{subsec:units}

Throughout the paper, we use units of \emph{physical} kpc and solar masses ($\mathrm{\mathrm{M}_{\odot}}$) - where not specified otherwise. We thus convert the simulation values from comoving kpc over $h$ to physical kpc at the redshifts of the two lenses, and the same units are adopted for the observational analysis. TNG50-1 has a softening length of $\epsilon_\mathrm{DM,*}=0.288$ kpc at $z=0$ (or comoving kpc), which we convert to physical kpc at the redshifts of the two considered systems.

\section{Observational data} \label{sec:lens_model}
\begin{table*}
    \centering
    \begin{tabular}{c|c|c|c|c|c|c}
        \hline
    \textbf{Lens} & \textbf{Profile} & $\Delta\log$E & $M_\mathrm{sub} [\mathrm{M}_{\odot}]$  &$c_\mathrm{sub}$ & $\kappa_\mathrm{sub} ["]$ & $\gamma_\mathrm{sub}$ \\
    \hline
                & PJ& 165 & $1.244\pm0.23\times10^{8}$  & - & - & -\\
      B1938+666  & NFW free & 169 & $2.026\pm0.584\times10^{8}$ & $256\pm140$ & -& -\\
        & NFW fix & 151 & $3.646\pm1.22\times10^{9}$  & fixed (8.93$^{+0.16}_{-0.2}$)& -& -\\
        & PL & 165 & -  & - & $0.0008\pm0.007$ & $2.42\pm0.19$\\
        \hline
        & PJ & 66  & $4.33\pm0.38\times10^{9}$  & - & - & -\\
       J0946+1006 & NFW free& 78 & $2.23\pm0.34\times10^{10}$ & $201\pm38$ &- & - \\
        & NFW fix& 15 & $1.09\pm0.5\times10^{12}$  & fixed (8.9$^{+0.64}_{-1.1}$) & - & -\\
        & PL & 83 & -  & - &$0.094\pm0.023$ & $2.11\pm0.11$\\
            \hline

    \end{tabular}

    \caption{Parameters of the detected subhaloes: we report the outcome of the nested sampling exploration performed with \textsc{MultiNest} for each lens and density profile model, in terms of mean values and the 95 per cent confidence interval. The corresponding posterior distributions are shown in Fig. \ref{fig:b1938}, \ref{fig:j0946} and \ref{fig:power-law}. The significance of the detection is expressed by the difference in Bayesian Evidence to the smooth model (third column): a positive difference indicates that the perturbed model is preferred. The PJ profile is defined by the subhalo mass $M_\mathrm{sub}$, which also characterises the NFW profile together with the subhalo concentration $c_\mathrm{sub}$. In the \emph{NFW fix} case, the concentration is determined by the concentration-mass-redshift relation from \citet{duffy08}, and the value reported here corresponds to the mean inferred $M_\mathrm{sub}$ while the error is propagated from the uncertainty on the mass. The PL profile is instead expressed in terms of normalised convergence $\kappa_\mathrm{sub}$ and the slope of the profile $\gamma_\mathrm{sub}$. }
    \label{tab:models}
\end{table*}

\begin{figure*}
\centering
\includegraphics[width=1\textwidth]{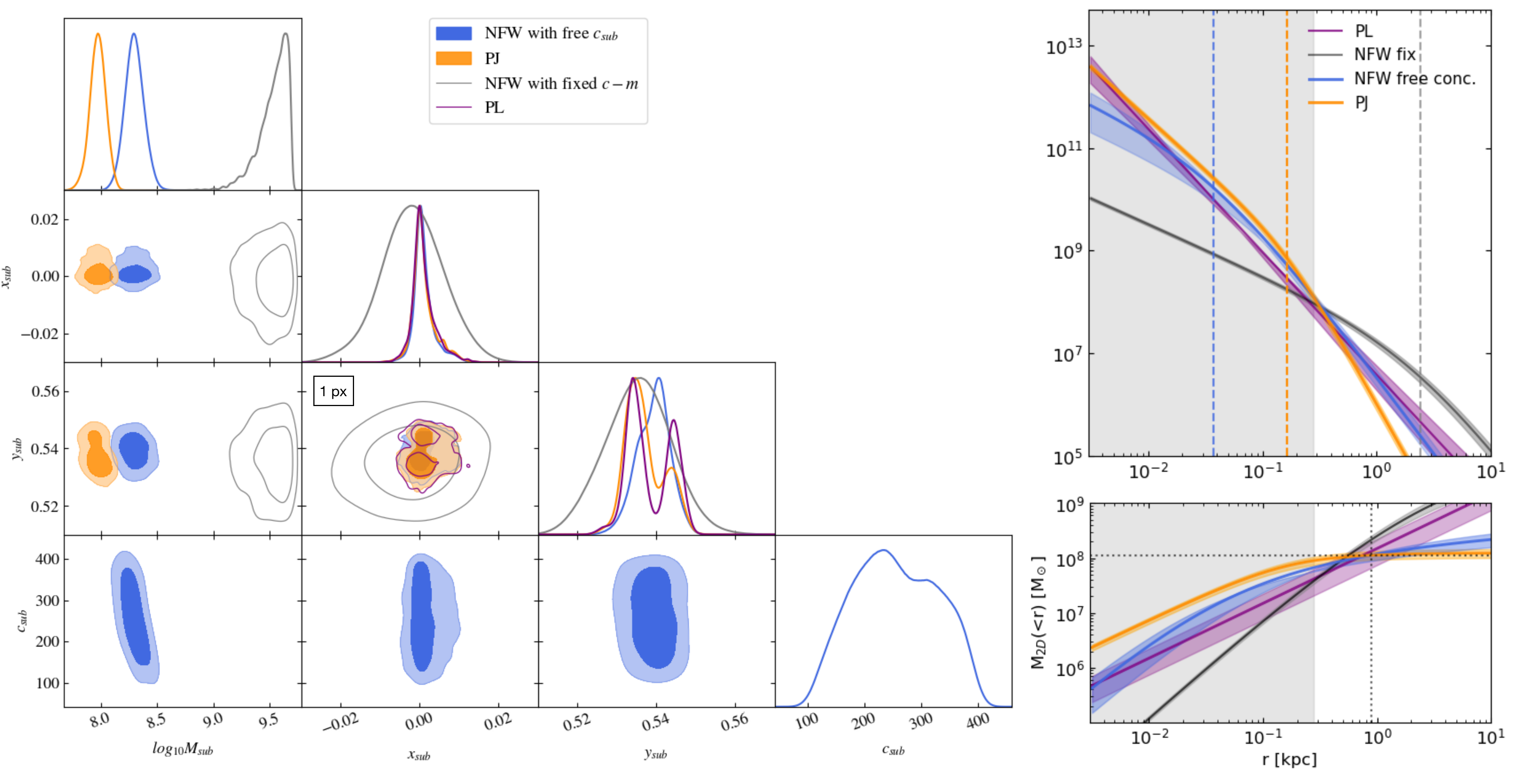}
\caption{ \emph{Left}: nested sampling posterior distributions for parameters describing the subhalo in the system B1938+666 for the PJ (orange) and NFW (blue and gray) profiles. In all cases, we fit for the subhalo mass $M_\mathrm{sub}$ (in $\mathrm{M}_{\odot}$) and position ($x_\mathrm{sub}$ and $y_\mathrm{sub}$, measured in arcseconds with respect to the centre of the image). We also plot the posterior of the subhalo position of the PL profile (purple), while the other parameters describing the models are plotted separately in Fig. \ref{fig:power-law}. We consider two variations of the NFW profile: one (grey) where the concentration $c_\mathrm{sub}$ is fixed by the concentration-mass relation from \citet{duffy08} and a case (blue) where $c_\mathrm{sub}$ is left free to vary. The best-fit values derived from the posteriors are listed in Table \ref{tab:models}. Note that we plot the mass $M_\mathrm{sub}$ in logarithmic space for better comparing the model, and this may give the impression that the grey contours are reaching the edge of the prior, which is not the case. The black rectangle in the ($x_{\rm sub}$-$y_{\rm sub}$) panel indicates the pixel scale of the data. The best lens model and reconstructed source for the smooth and smooth+NFW cases are shown in the Appendix in Fig. \ref{fig:lensing_data1}, and Fig. \ref{fig:lensing_data3} shows the posterior distribution for all parameters in all considered models. \emph{Right}: radial density and projected enclosed mass profiles corresponding to the best-fit subhalo profiles in the left panel, i.e. to the values reported in Table \ref{tab:models} (mean and 95 CL). The profiles are calculated analytically following the definitions in Sect. \ref{subsec:profiles}. The dashed vertical lines in the top panel mark the slope transition for the NFW and PJ profiles, i.e. $r_\mathrm{s}$ and $r_\mathrm{t}$. In the bottom panel, the dotted lines mark the distance at which the PJ, NFW and PL curves intersect, thus predicting the same mass, i.e. $r_\mathrm{eq}=0.9$ kpc. The grey areas show the data resolution.}
\label{fig:b1938}
\end{figure*}
We analyse two strong gravitational lensing systems where subhalo detections have been claimed in previous works. A summary of previous analyses of both systems can be found in Table \ref{tab:lenses}, and the optical/NIR images are shown in Fig. \ref{fig:lensing_data}. Here we use the same observations used by \citet{vegetti10} and \citet{vegetti12}, but our analysis has a few differences: we model the lens light and the lensed arc at the same time - without the need to subtract the former from the image - and we use a noise map which includes both the background and Poisson noise in each pixel.

\subsection{JVAS B1938+666}

The B1938+666 system was observed as part of the Jodrell Bank VLA Astrometric Survey \citep[JVAS;][]{jvas1}, and discovered to be a gravitational lens based on near-infrared HST imaging \citep{king97}.  In addition to the original NICMOS/NIC1 F160W HST observations (PropID 7255: PI Jackson), the lens system was observed by the WFPC2 and NICMOS/NIC2 instruments on HST in the F555W, F814W, and F160W bands (PropID 7495: PI Falco). The lens and source redshifts are $z_\mathrm{L}=0.881$ and $z_\mathrm{s}=2.059$ \citep{tonry_kochanek_2000,riechers2011}.  
In this paper, we use higher angular resolution observations that were obtained using the adaptive optics (AO) system at the Keck Telescope.  These K$^\prime$ data were obtained with the NIRC2 instrument as part of the SHARP Phase 1 sample and were used for both smooth mass models of the lensing galaxy \citep{lagattuta12} and the analysis by \citet{vegetti12}, who first discovered the presence of a dark perturber at the location of the brightest part of the northern arc, modelling it with a PJ profile and obtaining a subhalo mass $M_\mathrm{sub}=1.8\times10^{8}$ M$_{\odot}$. To date, this is the lowest mass of (sub)halo detected via gravitational imaging. They calculated 3$\sigma$ limits in magnitude and luminosity of $M_\mathrm{V}>-14.5$ and $L_\mathrm{V}<5.4\times 10^{7}$ L$_\mathrm{V,\odot}$. Despite the lower subhalo mass, the limits on its luminosity are not very stringent because the perturbation is detected in a luminous part of the arc, and it is thus not possible to distinguish its potential emission from the arc. As reported in \citet{vegetti12}, the limit was estimated by calculating the quadrature sum of the pixel noise in a 16x16 pixel square aperture, chosen to match the tidal radius of the substructure, and then applying a K-correction to convert the K$^\prime$ value in a rest-frame $V$ band limit. 

\citet{sengul22} analysed the lower-resolution HST observations of this system and measured the slope of the subhalo density profile, finding a best-fit value of $\gamma_{\rm sub}=1.96\pm0.12$, which corresponds to a concentration of $\sim60$. The inner slope is thus close to isothermal, and the concentration is high compared to the theoretical prediction for this mass range.
They also considered the possibility that the perturbation is caused by an isolated dark halo along the line of sight. They included the redshift as an additional free parameter and found a preferred redshift of $z=1.402^{+0.095}_{-0.154}$, thus preferring a line-of-sight halo over a subhalo. Here, we do not explore this possibility since we are interested in comparing different profile models with simulated subhaloes in hydrodynamical simulations and thus leave such considerations for a follow-up project.

\subsection{SDSS J0946+1006}
We use observations of the gravitational lens system J0946+1006 observed by the ACS camera on board the HST as part of the SLACS sample \citep{bolton06}. Observations have revealed the presence of three lensed sources: two arcs visible in HST observations \citep{gavazzi08} $z_\mathrm{s1} = 0.609$ (inner lensed arc) and $z_\mathrm{s2} = 2.035$ (outer lensed arc) and an additional higher-redshift source at $z_\mathrm{s3} = 5.975$ detected with \textsc{MUSE} \citep{collett20}. In this work, we consider observations in the F814W band and model the inner arc, which also is the one with the highest signal-to-noise ratio.
The detection of a dark perturber of mass $M_\mathrm{sub}=3.51\times10^{9}$ M$_\odot$ in the north-west part of this arc has been claimed first by \citet{vegetti10}. They also calculated a 3$\sigma$ upper limit luminosity of the subhalo of  $L_\mathrm{V}<5.0\times 10^{6}$ L$_\mathrm{V,\odot}$. Based on the residual images, they determined an upper limit on the magnitude of the substructure in two different ways: by setting the limit equal to three times the estimated (cumulative) noise level or by aperture-flux fitting, both inside a 5x5 pixel window. The aperture is chosen to gather most of the light of the substructure, which is expected to be effectively point-like given its low mass. Its presence and location were later confirmed by other works \citep{minor2021,nightingale24,ballard24}, while the estimates of its total mass and properties depend on the assumptions made for the subhalo profile (see Table \ref{tab:lenses}). The original detection has been modelled with a Pseudo-Jaffe \citep{jaffe83,munoz01} profile that produces a very compact mass distribution and is thus very efficient as a lens. Other works instead described the subhaloes as NFW profiles (classic or truncated). Due to the shallower inner slope, these require a higher total mass to produce a lensing effect of the same magnitude. Indeed, \citet{nightingale24} finds a total subhalo mass $\sim10^{11} \mathrm{M}_\odot$ by calculating $M_\mathrm{sub}^{200}$ and assigning it a concentration $c_{200}$ following \citet{ludlow16}. Their findings are compatible with the predictions from \citet{despali18}: the mass of an NFW profile (following a concentration-mass relation) must be one order of magnitude larger than a PJ subhalo to produce a lensing effect of the same scale. \citet{minor2021} and \citet{ballard24} have instead considered the concentration as an additional free parameter in the NFW profile, finding best-fit values that are extreme outliers of the CDM model.  
 
\section{Results of lens modelling} \label{sec:results}

We model the two considered systems with the \textsc{pronto} code (Vegetti et al. in prep), as described in Sect. \ref{subsec:code}. We first model the main lens, including both the central galaxy's mass and light, finding lens parameters consistent with previous works. We then add a second mass component to describe the perturber and test the density profiles listed in Sect. \ref{subsec:profiles}. For each case, we calculate the posterior distribution of the lens and subhalo parameters simultaneously using \textsc{MultiNest} \citep{feroz13}. In this section, we focus on the posterior distributions of the subhalo parameters and their physical interpretation while we show the full posteriors, the best lens models and reconstructed sources in the Appendix (Fig. \ref{fig:lensing_data1}-\ref{fig:lensing_data4}). 

The best-fit parameters describing the subhaloes and the associated errors are listed in Table \ref{tab:models}: these are the mean values and 95 confidence intervals calculated by \textsc{MultiNest}. It is a well-known fact that these errors are somewhat underestimated and that \textsc{MultiNest} returns overly optimistic uncertainties underpredicting the size of the posteriors. For this reason, we follow \citet{rizzo18} and provide more realistic uncertainties by summing in quadrature the errors from \textsc{MultiNest} and the difference between the MAP and mean values of the posterior distribution. The third column reports the logarithmic difference in Bayesian Evidence between the perturbed and smooth model $\Delta\log E = \log  E_{\rm pert} -  \log E_{\rm smooth}$: a positive difference indicates that the perturbed model is preferred over the smooth case. 
In most cases, the additional mass component significantly improves the Bayesian Evidence of the model. The condition $\Delta\log E>50$ \citep{vegetti14,ritondale19b} condition is thus satisfied, thus confirming the detections. 

\subsection{Inferred properties of the detected subhaloes} \label{subsec:lensmodels}

\begin{figure}
\centering
\includegraphics[width=\columnwidth]{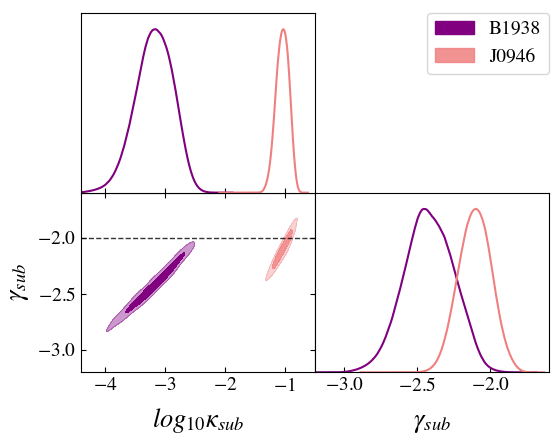}
\caption{Posterior distribution of the Power-Law model parameters: the slope $\gamma_\mathrm{sub}$ and the convergence normalisation $\log(\kappa_\mathrm{sub})$, measured in arcseconds. The dashed lines mark the slope of an isothermal profile. }
\label{fig:power-law}
\end{figure}

Figures \ref{fig:b1938}, \ref{fig:j0946} and \ref{fig:power-law} report the posterior distributions of the subhalo properties for the PJ, NFW and PL profiles. The right panels of Figures \ref{fig:b1938} and \ref{fig:j0946} show the analytic density and enclosed mass profiles corresponding to the best-fit subhalo parameters; we discuss these in detail in Sect. \ref{subsec:encmass}.

In the case of B1938+666, all four profiles significantly improve the Bayesian Evidence compared to the smooth model (see Table \ref{tab:models}), with the NFW profile with fixed concentration being the least favourite model. The inferred subhalo positions, left free to vary, are consistent within the resolution of the data (70 \emph{mas}), indicating the stability of the detection. The position is consistent with \citet{vegetti10}. For J0946+1006, the PJ, PL and NFW profile with free concentration also improve the Bayesian Evidence of the model, while NFW profile with fixed concentration improves the evidence compared to the smooth model but does not fit the data equally well (i.e. does not satisfy the condition for a reliable detection, $\Delta\log E<50$). We note how the inferred subhalo position for this model also differs by more than 2$\sigma$ from the other profiles. In this case, the inferred subhalo mass is extremely high, and it cannot correspond to a dark structure. This result is compatible with the highest mass value (and the corresponding subhalo position) calculated by \citet{nightingale24}. In practice, such a high mass cannot lie very close to the arc because it would modify the lensed images far from the detection location and is thus forced to move (see Fig. \ref{fig:lensing_data2b} for the results of the fit with different NFW profiles). For the other three profiles, the subhalo position is instead consistent within the resolution (90 \emph{mas}) and with \cite{vegetti12}.

While PJ and NFW models have fixed values for the inner and outer density profile slopes, when we model the perturbers with a (not truncated) power-law profile, the slope is left as a free parameter. In practice, this is defined as in equation (\ref{eq:lens}), but assuming spherical symmetry ($q=1$). Fig. \ref{fig:power-law} shows the resulting posteriors for the subhalo convergence normalisation $\kappa_{0}$ and inner slope $\gamma_\mathrm{sub}$. In both lenses, the subhalo position agrees with the results of the PJ and NFW profiles (see also Fig. \ref{fig:zoom} and the Appendix). The extremely compact detection in B1938+666 is best described by a profile with $\gamma_\mathrm{sub}=-2.4\pm0.17$, steeper than isothermal and than all profiles considered so far. The inferred slope in J0946+1006 is instead $\gamma_\mathrm{sub}=-2.1\pm0.11$, which is still compatible with the isothermal slope and thus with the PJ and mNFW profiles. 

\begin{figure*}
\centering
\includegraphics[width=0.35\textwidth]{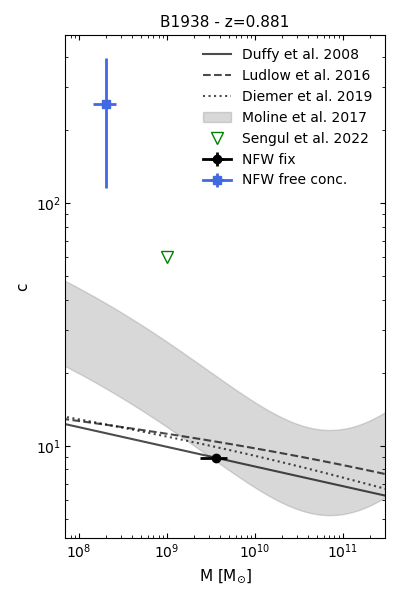}
\includegraphics[width=0.35\textwidth]{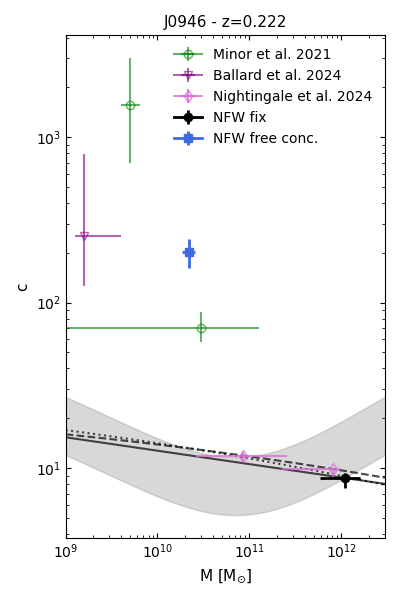}
\caption{Comparison of the NFW models from our analysis (black circles and blue squares) and previous works with concentration-mass-redshift relations derived from N-body simulations. All results are expressed in terms of the virial mass of haloes $M_{\rm vir}$, except for \citet{moline17} who measured the relation using $M_{\rm 200}$, i.e. the mass within the radius enclosing an overdensity of 200. In this case, the relation has an additional dependence on the distance of the subhalo from the centre of its host halo. Since the observational results are derived in projection (on the lens plane), we cannot determine the three-dimensional distance from the host centre, and we plot a range of relations for distances between 0.1$R_{\rm vir}$ and $R_{\rm vir}$. Given that it was calibrated on subhaloes rather than isolated haloes, it naturally predicts higher concentrations at the low-mass end. In each panel, the coloured symbols and errorbars show our inferred values from the analysis of the observations (black circle and blue square) and the results from previous works (empty symbols), listed in Table \ref{tab:models}. \citet{sengul22} used a fixed mass $M_{\rm sub}=10^{9}$ M$_{\odot}$ and varied the inner slope of the profile (that we report in the Table), from which they estimated a corresponding concentration but did not provide uncertainties in this quantity.}
\label{fig:concentration}
\end{figure*}

Given the intrinsic difference among the adopted profiles and mass definitions, the inferred subhalo mass differs, and our measurements are consistent with the trends in previous works, reported in Table \ref{tab:lenses}.
When the concentration is left free to vary, the resulting best-fit NFW profiles are very compact, leading to large concentrations, consistent with previous works:  $c_{\rm sub}=256\pm140$ in B1938+666 and $c_\mathrm{sub}=201\pm38$ in J0946+1006. In practice, this means that the classic NFW profile is not a good description of the properties of the detections: to reproduce them, the scale radius $r_\mathrm{s}$ becomes artificially small, leading to extreme values of concentration. To aid the comparison with theoretical predictions, in Fig. \ref{fig:concentration}, we plot the values inferred from the observational analysis together with the concentration-mass relations measured in numerical simulations \citep{duffy08,ludlow16,diemer19,moline17}. By construction, the results obtained with a fixed concentration-mass relation - our black points and the results from \citet{nightingale24} - lie on the analytical predictions, while all the other measurements show a clear departure from the numerical predictions. We note, however, that all these concentration-mass relations result from fitting the dark matter profiles of haloes in simulations that do not include the effect of baryons, which may be especially relevant at the mass scales corresponding to the detection in J0946+1006. \citet{mastromarino23} measured the concentration-mass relation in the Eagle cold and self-interacting dark matter runs \citep{robertson21}, comparing the dark-matter-only and hydro values. At the scale of galaxies with masses M$_{200}>10^{12}M_{\odot}$, the hydro concentrations are higher by a factor of 2-3 compared to the dark-matter-only case. It is not possible to extrapolate their results to the mass range considered here. Still, their results suggest that the presence of baryons would not increase the \emph{mean} concentration enough to reproduce our values inferred from observations.

In Sect. \ref{sec:comparison}, we compare our results to the subhaloes of TNG50-1, which combines high-resolution with advanced treatment of baryonic physics. 

\subsection{A radius of equal enclosed mass} \label{subsec:encmass}

\begin{figure*}
\centering
\includegraphics[width=\textwidth]{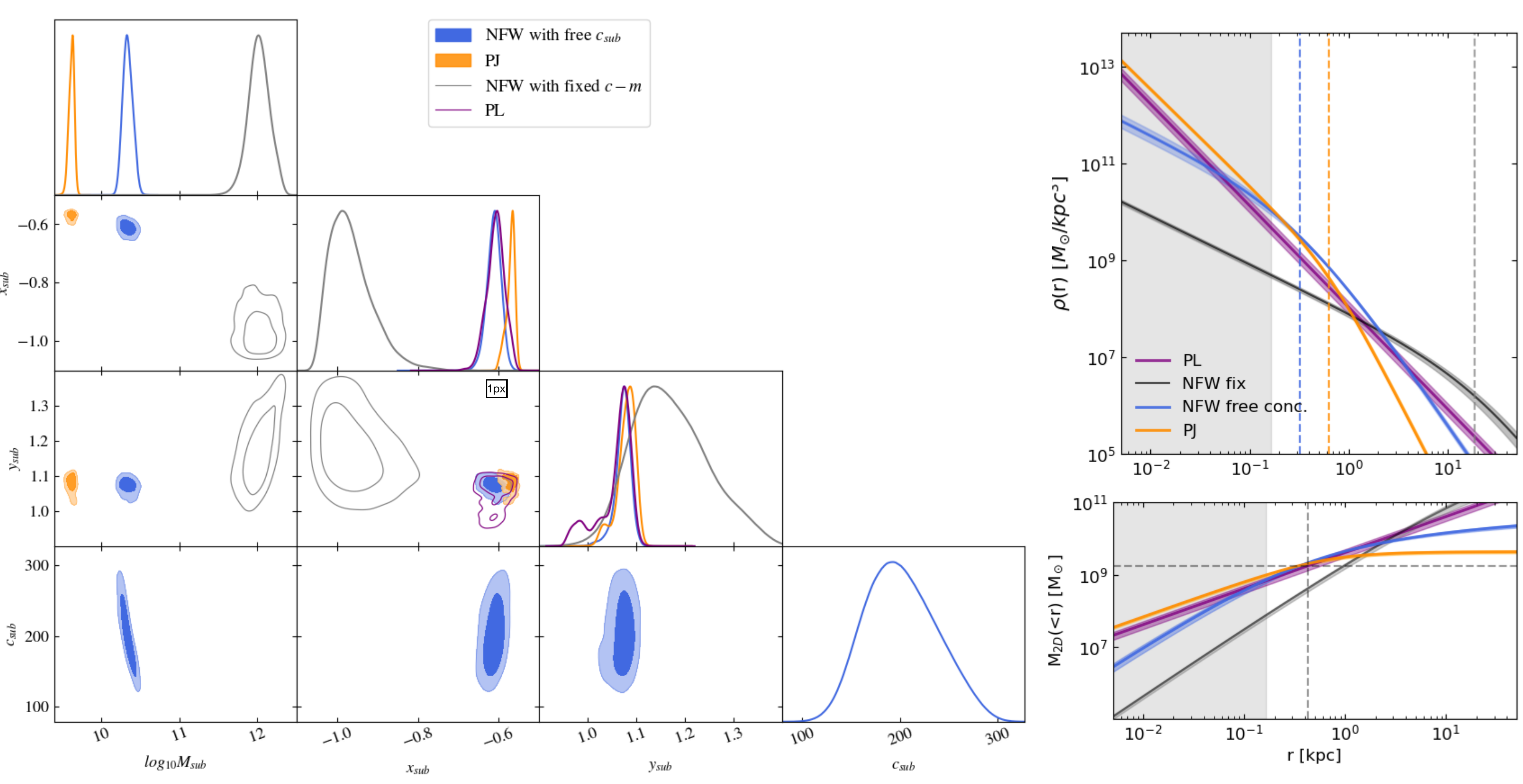}
\caption{Same as Fig. \ref{fig:b1938} but for the lens system J0946+1006.  Here, the NFW profile with concentration fixed by the concentration-mass relation is a worse fit to the data than the other models: a symptom is the best-fit position of the subhalo, which is very different and more uncertain than for the other profiles. Moreover, the enclosed mass in the bottom-right panel does not converge within a radius similar to the other two profiles, i.e. $R_\mathrm{eq}=0.45$ kpc. The best models and the full posterior distributions can be found in the Appendix in Figures \ref{fig:lensing_data1} and \ref{fig:lensing_data4}.}
\label{fig:j0946}
\end{figure*}

Interestingly, the PJ, PL and NFW (with free concentration) profiles are preferred over the smooth model at roughly the same level (see Table \ref{tab:models}), meaning that they can fit the images equally well despite their structural differences. To investigate the reason behind this, we look at their density and enclosed mass profiles, plotted in the right panels of Figures \ref{fig:b1938} and \ref{fig:j0946}. The grey bands mark the resolution of the data, calculated as the FWHM of the PSF (divided by 2, given that we are looking at a radial profile). For the PJ and NFW profiles, the steeper inner part of the density profiles is thus close to the resolution of the data in both cases: in B1938+666, both $r_\mathrm{t}$ and $r_\mathrm{s}$ are unresolved, while in J0946+1006, they lie right above the resolution limit. This means that the lensed images can resolve the profiles in the slope transition region and above but not in the innermost part. The untruncated PL profiles produce an inner slope between the two models. 
In the bottom panels, we plot the projected enclosed mass $M_\mathrm{2D}(<R)$: this is especially important since gravitational lensing robustly measures the projected mass distribution - which we then convert to a 3D profile. For all models that cause an increase in the Bayesian Evidence of the model of $\Delta\log E>50$, the enclosed mass profiles intersect at roughly the same distance $R_\mathrm{eq}$ from the subhalo centre: $R_\mathrm{eq}=0.9$ kpc for B1938+666 and $R_\mathrm{eq}=0.45$ kpc for J0946+1006. Note that $R_\mathrm{eq}$ encloses almost the total mass of PJ profiles, while the NFW and PL have a larger fraction of the mass in the outer parts. A counter-example is the NFW profile with a fixed mass-concentration relation in J0946+1006, which is a worse fit and does not converge to the same enclosed mass within the same radius. We can conclude that all subhalo models are trying to fit the enclosed mass within $R_\mathrm{eq}$ rather than the inner slope, which is not resolved.

\begin{figure*}
\centering
\includegraphics[height=0.3\textwidth]{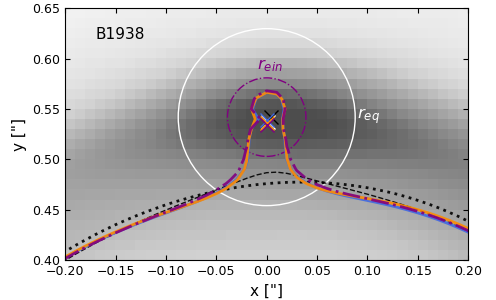}
\includegraphics[height=0.3\textwidth]{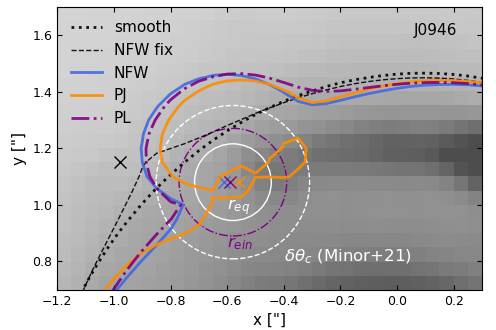}
\caption{We zoom on the location of the two detections to highlight the differences in the critical lines and subhalo position between the considered models (lines and crosses of the corresponding colour). The circles marks $(i)$ the radius $R_\mathrm{eq}$ at which the PJ and NFW profiles produce the same enclosed mass (solid white), $(ii)$ the Einstein radius $R_\mathrm{ein}$ calculated with equation (\ref{eq:rein}) for the PL profile (purple), $(iii)$ the size corresponding to the \emph{robust mass} defined for J0946+1006 by \citet[][dashed white]{minor2021}. In the case of B1938+666, the perturber is very compact, and the three best models have very similar critical lines. In J0946+1006, the larger subhalo mass translates into larger visible differences among the models.  }
\label{fig:zoom}
\end{figure*}

To investigate the nature of $R_{\rm eq}$, in Fig. \ref{fig:zoom} we visually compare all models in the lens plane, looking at their impact on the critical lines around the location of the perturbations. Each model is represented by a different colour (that matches Figures \ref{fig:b1938} and \ref{fig:j0946}) and line style, and the inferred subhalo position is marked by crosses of the corresponding colour. The white circles show $R_{\rm eq}$. Moreover, we used equation (\ref{eq:rein}) to calculate the Einstein radii of the PL model, represented by the purple dashed circle. In the case of B1938+666 (left panel), the three best-fitting profiles produce an extremely similar effect that is needed to fit the very compact mass distribution of the subhalo, while the NFW model with fixed concentration barely modifies the critical lines with respect to the smooth model. 
The perturber in J0946+1006 (right panel) is more massive and thus produces a larger variation of the critical lines, with a visible difference between the models. Here, we also compare to the results by \citet{minor2021} who modelled the same system: with their model, they inferred a \emph{robust radius} $\delta\theta_{c}=1 \ \mathrm{kpc}$, defined by \citet{minor17} as the radius within which the mass measurements is robust. We plot it here as the white dashed circle - assuming as a centre our best-fits subhalo position (which may differ from their original result). Interestingly, the three radii plotted here do not match: it is intriguing that $R_\mathrm{eq}$ does not yet correspond to an obvious quantity related to gravitational lensing theory. However, $\delta\theta_{c}$ also depends on the macro model and thus, part of the difference between their value and ours could be due to the difference in the underlying mass distribution of the lens. It is reasonable that detectable perturbations would robustly characterise the projected surface mass density at their location. A more systematic analysis is needed to identify its nature, given that it would allow us to interpret the detections without the need to impose a density profile model. We investigate this scale in a follow-up paper (Tajalli et al. in prep).

\section{Comparing observations and simulations} \label{sec:comparison}

\begin{figure}
\centering
\includegraphics[width=\columnwidth]{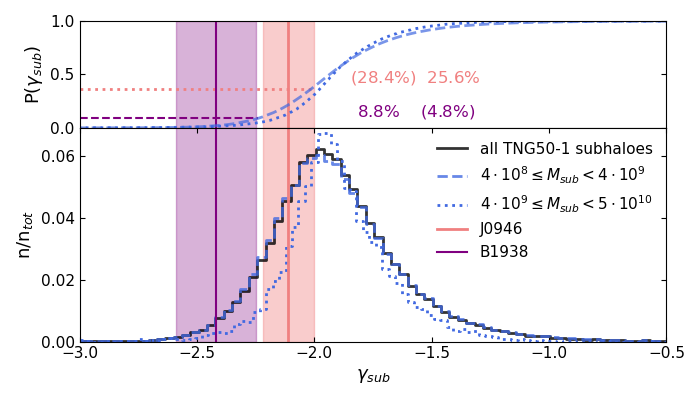}
\caption{Comparison between the $\gamma_{\rm sub}$ inferred from observations with the PL model (see Fig. \ref{fig:power-law}) and the inner slopes of subhaloes from the TNG50-1 simulation. In the latter, the slope has been measured for all subhaloes in \citetalias{heinze2023} (see their Fig. 9). The bottom panel shows the distribution of slopes of all subhaloes (black) and two mass bins roughly corresponding to the range of the detections. The coloured vertical lines and shaded areas correspond to the observed values (Table \ref{tab:models}). In the top panel, we show the cumulative distribution for all subhaloes and the second mass bin, and we report the corresponding fractions of subhaloes that are compatible with the observed slopes in the two mass bins, highlighting the one that better corresponds to the detection in each case with the horizontal lines and corresponding numbers.}
\label{fig:power-law2}
\end{figure}

\begin{figure*} 
\centering
\includegraphics[width=\textwidth]{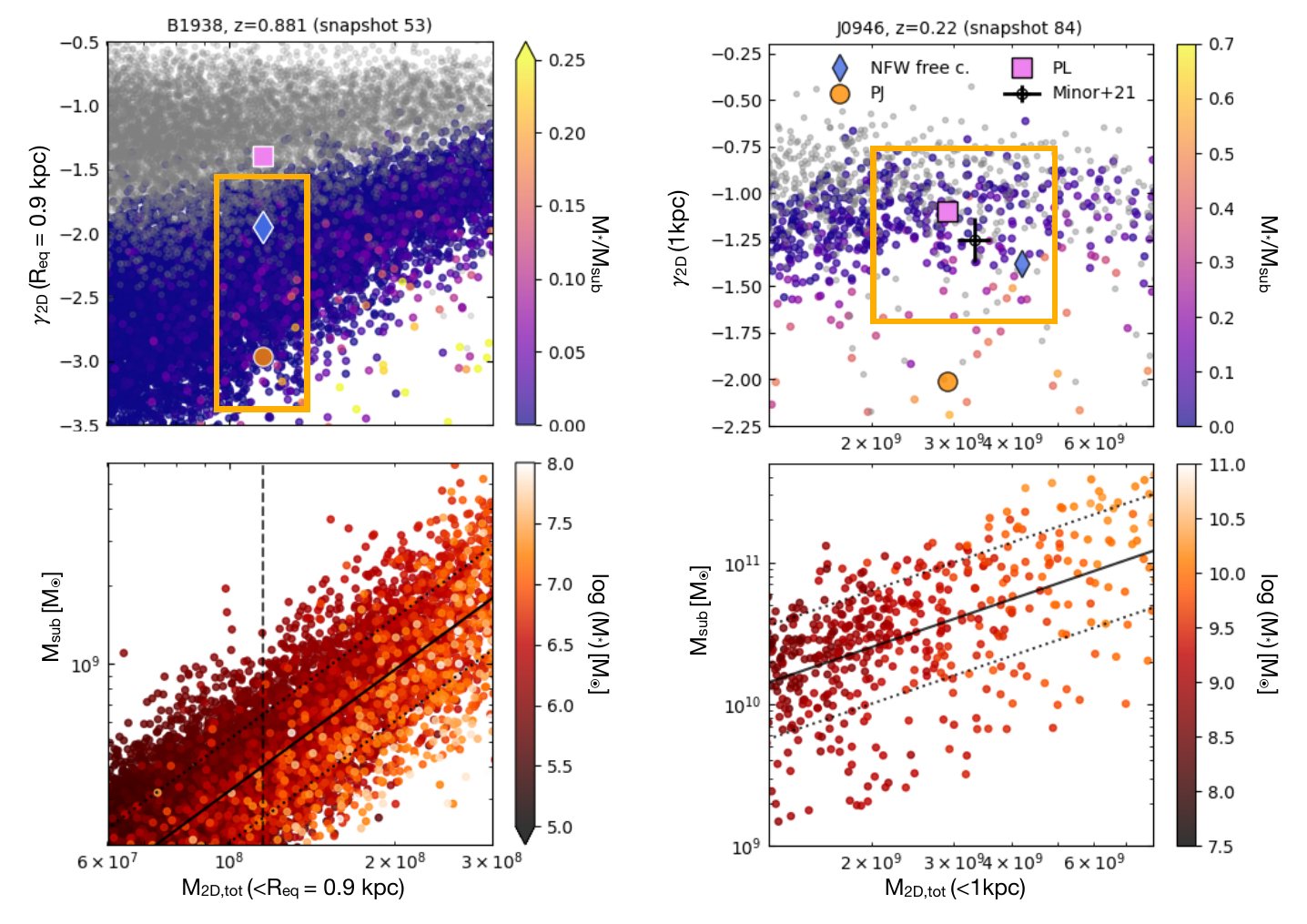}
\caption{\emph{Top:} we plot log-slope of the surface density profile around  $r=R_{\rm eq}=0.9$ kpc (for B1938+666) or  $r= 1$ kpc (for J0946+1006) against the projected $total$ mass enclosed within the same distance. See Fig. \ref{fig:resolution} for the J0946+1006 results at $r=R_{\rm eq}$, which we believe are affected by resolution. The coloured points indicate the top 10 per cent of lines of sight that yielded the largest density values, colour-coded by the stellar-to-total mass ratio. The grey dots indicate the averages over all 1000 lines of sight. The larger points show the slope and mass values inferred from our lens modelling for the PJ (orange circle), NFW with free concentration (blue diamond) and PL (pink square) profiles. In the right panel, we also compare to the measurement by \citet{minor2021}. \emph{Bottom:} we plot the total subhalo mass as a function of the enclosed projected mass. Points are coloured by the total stellar mass of the subhalo. 
The mean and standard deviation used in equation (\ref{eq:fit_m2d}) are represented by the solid and dotted black lines.
}
\label{fig:gamma_2d_vs_m2d_1}
\end{figure*}

\begin{figure*} 
\centering
\includegraphics[width=\columnwidth]{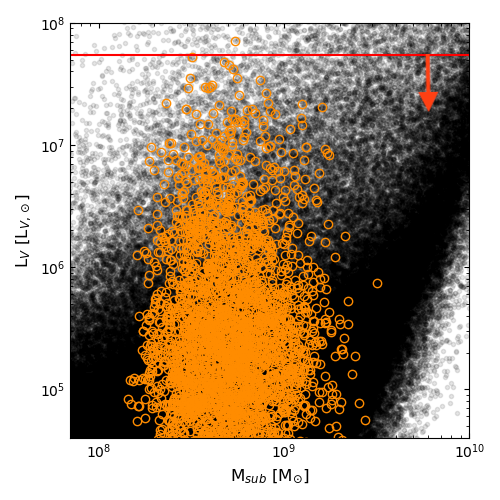}
\includegraphics[width=\columnwidth]{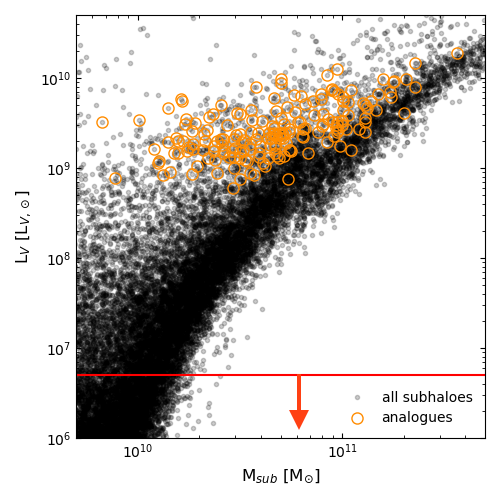}
\caption{We compare the luminosity of the simulated subhaloes to the observational limits set for the two detections. First, we select the systems that best match the observational inference in terms of projected slope and enclosed mass: in practice, these are selected within the orange rectangles in the top panels of Fig. \ref{fig:gamma_2d_vs_m2d_1}. We compute their luminosity L$_{V}$ from the V-band magnitudes provided in the TNG50-1 subhalo catalogue and plot them as orange circles as a function of the subhalo mass M$_{\rm sub}$. We show the results for B1938+666 on the left and J0946+1006 on the right. The black dots show instead the luminosity of the general subhalo population at the same redshift. The observational upper limits (see Sect. \ref{sec:lens_model}) are represented by the red horizontal line in the plot.}
\label{fig:luminosity}
\end{figure*}

\begin{figure*} 
\centering
\includegraphics[width=0.9\textwidth]{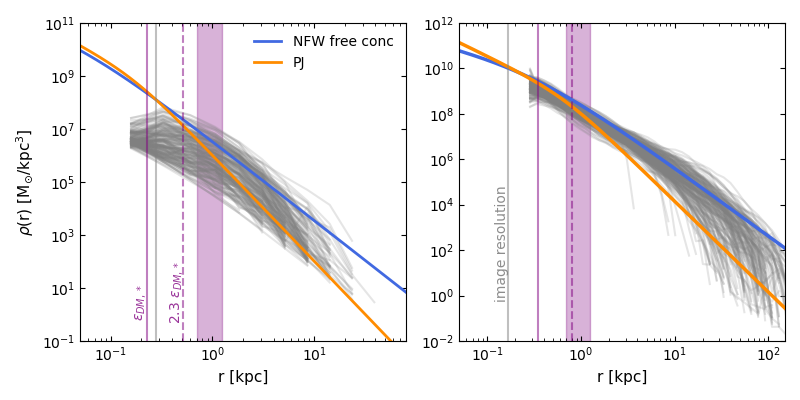}
\caption{Subhalo density profiles of the simulated subhaloes that best match the observational inference (B1938+666 on the left and J0946+1006 on the right), i.e. the same shown in orange in Fig. \ref{fig:luminosity}. The purple band shows the radial range used to measure the slope $\gamma_\mathrm{2D}$ for Fig. \ref{fig:gamma_2d_vs_m2d_1}. The vertical lines mark the softening of the simulation (solid purple), and the value commonly used in simulation as a reliable resolution limit (i.e. $2.3\epsilon_\mathrm{DM,*}$, dashed purple) and the resolution of the observational data. }
\label{fig:prof}
\end{figure*}

In this section, we use the TNG50-1 simulation to compare simulated subhaloes with the results of the previous section. \citetalias{heinze2023} already showed that the classic NFW model does not provide a good description of subhalo density profiles in TNG50 because the measured inner logarithmic slopes are steeper, and the NFW cannot model the external truncation. This is due to the combined effects of baryons at the centre of satellites and haloes and tidal stripping.
In this section, we search for simulated analogues of the two detections, compare to the results of \citet{minor2021} and discuss their properties in terms of density profiles, projected mass, stellar content, merger history and lensing effect. 

\cite{minor2021} used the TNG100-1 simulation, while we can count on the higher-resolution TNG50-1 run: the spatial resolution, defined by the gravitational softening, improves from $\epsilon_\mathrm{DM,*}=0.74$ comoving kpc to $\epsilon_\mathrm{DM,*}=0.28$ comoving kpc; similarly, the dark matter particle masses decreases from $m_\mathrm{DM}=7.5\times10^{6}$ M$_{\odot}$ to $m_\mathrm{DM}=4.5\times10^{5}$ M$_{\odot}$. This provides us with a better resolution at the centre of haloes, and we are able to measure density profiles down to lower halo masses. However, all simulations are prone to numerical effects, and we discuss our limitations in Sect. \ref{subsec:resolution}. We choose the simulation snapshots that are closest to the lens redshift $z_\mathrm{L}$ of each lens: snapshot 84 at $z=0.2$ and snapshot 53 at $z=0.89$.

\subsection{Finding analogues of the detections in CDM}  \label{subsec:analogues}

We start by comparing the inner slopes inferred from the data with the PL model (see Fig. \ref{fig:power-law}) to those of simulated subhaloes. We use the inner slopes calculated in \citetalias{heinze2023} by fitting a \emph{modified Schechter profile} to the simulated data, i.e. a PL profile with exponential truncation. The two panels of Fig. \ref{fig:power-law2} show the cumulative and differential distribution of subhalo slopes in different mass bins. Here, we use all TNG50 subhaloes, independently of their host halo mass, to determine how common the inferred slope is in the entire population. Additionally, we choose two bins corresponding to the ranges of inferred masses for the two detections (see Table \ref{tab:models}). The distribution for subhaloes with $M_\mathrm{sub}<4\times10^{9}$ M$_{\odot}$ is essentially identical to that of the entire population, while the distribution for the higher mass bin moves to shallower slopes. When looking only at density slopes, the detection in J0946+1006 is thus compatible with the slopes of 25.6 per cent of subhaloes in the range $4\times10^{9}$ M$_{\odot}\leq M_\mathrm{sub}<5\times10^{10}$ M$_{\odot}$. The perturber in B1938+666 is instead more compact and has a steeper - and thus rarer - profile slope that is compatible with 8.8 per cent of subhaloes in the lower mass range $4\times10^{8}$ M$_{\odot}\leq M_\mathrm{sub}<4\times10^{9}$ M$_{\odot}$. 

However, this comparison is not yet complete since we have shown in Sect. \ref{subsec:lensmodels} that the subhalo profiles must match not only in terms of slope but mostly in the enclosed mass. Moreover, the histograms are binned using the total subhalo mass, while the most robust mass measurements through lensing correspond to the enclosed mass within the radius at which different profile definitions converge (see Figures \ref{fig:b1938} and \ref{fig:j0946}). 

We now proceed to a more thorough comparison in the parameter space of projected quantities. We adopt the definitions of \citet{minor2021}, who used $\gamma_\mathrm{2D}(R_{c})$ and $M_\mathrm{2D}(<R_{c})$, i.e. the average log-slope of the surface density profile around a characteristic radius and the projected mass within the same distance. In the case of B1938+666, we choose $R_{c}=R_{\rm eq}=0.9$ kpc, since this scale corresponds to $\sim5.9\epsilon_\mathrm{DM,*}$ at $z=0.881$ and thus is resolved reasonably well. In the case of J0946+1006, $R_{\rm eq}=0.45$ kpc = $\sim1.9\epsilon_\mathrm{DM,*}$  at $z=0.222$ which is not enough to guarantee good convergence, as we discuss in more detail in Sect. \ref{subsec:resolution}. In the latter case, we thus choose the same value adopted by \citet{minor2021}, $R_{c}=1$ kpc; this has the advantage of being both reliable in terms of resolution and comparable to their work. 

We calculate $\gamma_\mathrm{2D}(R_{c})$ and $M_\mathrm{2D}(<R_{c})$ from the PJ, NFW and PL analytical profiles that best fit the observational detections, i.e. those shown in Figures \ref{fig:b1938}, \ref{fig:j0946} and \ref{fig:power-law} and corresponding to the values in Table \ref{tab:models}. In the simulation, the projected mass within the chosen radius $M_\mathrm{2D}(<R_{c})$ is calculated by summing up the masses of all subhalo particles that lie within a cylinder with a radius $R_{c}$. The average log-slope around the characteristic radius $M_\mathrm{2D}(R_{c})$ is calculated by computing the surface density profile around $R_{c}$ using five bins consisting of cylindrical shells and fitting a power law to the data points (the considered range is shown by the purple band in the bottom panel of Fig. \ref{fig:gamma_2d_vs_m2d_1}). Both quantities are computed for 1000 random orientations and averaged. Moreover, we compute the average values for the top 10 per cent lines of sight with the highest values of $\gamma_\mathrm{2D}(R_{c})$ and $M_\mathrm{2D}(<R_{c})$ respectively.  \citet{minor2021} only considered subhaloes hosted by haloes of $M_\mathrm{halo}\sim 10^{13}\mathrm{M}_{\odot}$, i.e. the typical mass of lens galaxies. Given that the simulation provides us with a (50 Mpc)$^{3}$ volume instead of (100 Mpc)$^{3}$ - and thus fewer objects - we relax the selection criteria to include larger statistics and consider host haloes with total masses of $5\times10^{12} - 5\times10^{13}$ M$_\odot$ and stellar masses of $10^{10} - 10^{13}$ M$_\odot$. \citet{mastromarino23} showed that haloes selected in this range of viral masses have a distribution of Einstein radii similar to that of observed lens samples.

The top panel of Fig. \ref{fig:gamma_2d_vs_m2d_1} shows the comparison results in the parameter space defined by $\gamma_\mathrm{2D}(R_{c})$ and $M_\mathrm{2D}(<R_{c})$: the grey points show the averages taken over all lines of sight, while the top 10 per cent values are coloured by the fraction of stellar mass of each subhalo. On the left, we see that many subhaloes match the NFW profile that best fits the detection in B1938+666. The PJ and PL profiles are also compatible with simulated subhaloes, although they lie at the edge of the distribution. However, we must note that there is a large difference between the grey and coloured points, indicating that high-density projections are needed to reproduce the observed results. The subhaloes are dark matter dominated, i.e. with a fraction in stellar mass lower than $\sim 5$ per cent. In the right panel, we are able to find a few subhaloes matching the inner slope predicted by the PL and NFW profiles for the detection in J0946+1006. In this case, the fraction of stellar mass of the subhaloes is higher but still less than 15 per cent. These are close to the constraints from \citet{minor2021}, represented by the black cross. However, they did not find any good match when performing the same measurements on the TNG100-1 subhaloes, concluding that this detection is an outlier of CDM. As we discuss more in detail in Sect. \ref{subsec:resolution}, we believe that the difference in our results is explained by the better resolution of TNG50-1, which allows us to measure the inner density profiles of subhaloes more reliably. 

\subsection{Properties of the simulated analogues}

To highlight the range of subhalo properties that could match the detections, we now look at their total and stellar masses, density profiles, luminosities and merger histories.

In the bottom panels in Fig. \ref{fig:gamma_2d_vs_m2d_1}, we plot the total subhalo mass M$_{\rm sub}$ as a function of $M_\mathrm{2D}(<R_{c})$ and colour the points by their stellar mass. In both cases, a certain value of $M_\mathrm{2D}(<R_{c})$ corresponds to total subhalo masses that span one order of magnitude. Moreover, while the total stellar masses are low for the B1938+666 detection ($M_{*}\leq10^{6.5}$ M$_{\odot}$), those required to match the properties of J0946+1006 are higher ($10^{8.5}$ M$_{\odot}\leq M_{*}\leq10^{10}$ M$_{\odot}$) and thus may be problematic for a subhalo detection that does not correspond to a visible satellite. We fit the mean linear relation between the total and projected mass as 
\begin{equation}
    \log(M_\mathrm{sub})=A+B\log(M_\mathrm{2D}), \label{eq:fit_m2d}
\end{equation}
finding $(A,B) = (-3.95 \pm 0.2, 1.65)$ in the left panel and $(A,B)= (-0.08 \pm 0.4, 1.13)$ in the right one. These are plotted as mean black solid (mean) and dotted (standard deviation) lines.

We now select simulated subhaloes that lie close to the detections in the plane defined by $\gamma_\mathrm{2D}(R_{c})$ and $M_\mathrm{2D}(<R_{c})$ and investigate their properties more in detail. The selection of such \emph{analogues} is shown by the orange boxes in the top panels of Fig. \ref{fig:gamma_2d_vs_m2d_1}: for B1938+666, we choose subhaloes that match the whole range of slopes between the NFW and PJ results, given that they all correspond to dark-matter-dominated structures; for J0946+1006, we instead exclude systems that match the PJ slope, given that they are all baryon-dominated (M$_{*}$/M$_{\rm sub}$>0.5). First, we compare the luminosities of simulated subhaloes to the upper limits that have been estimated from observations: $L_\mathrm{V}<5.4\times 10^{7}$ L$_\mathrm{V,\odot}$ for B1938+666 and $L_\mathrm{V}<5\times 10^{6}$ L$_\mathrm{V,\odot}$ for J0946+1006. We use the absolute magnitudes in V band \citep{buser78} provided in the TNG50-1 subhalo catalogue based on the summed-up luminosities of all the stellar particles of each subhalo. The resulting values for the analogues are shown in Fig. \ref{fig:luminosity} as orange circles and compared to the general population of subhaloes (black dots). In both cases, the analogues of the detection have higher luminosities than the mean population, and this is particularly evident in the case of J0946+1006 (right panel). While the left panel shows that all analogues are compatible with the upper limit set for B1938+666, on the right, we see the opposite: all simulated analogues are too bright to correspond to the subhalo in J0946+1006. As discussed in Sect. \ref{sec:lens_model}, we point out that the upper limit is more stringent for J0946+1006, given that the detection is located in a relatively dark part of the image, while it could be not as accurate for B1938+666. Nevertheless, this confirms that the subhalo detection in J0946+1006 is in tension with the predictions from numerical simulations. These considerations may not hold if the detected object was an isolated halo along the line of sight and not a subhalo: the simulated luminosities may change, as well as the observational limits, since an object further away from the lens may be harder to see; moreover, the mass and concentration of an isolated halo may differ from the values calculated from subhaloes.

\begin{figure*} 
\centering
\includegraphics[width=\textwidth]{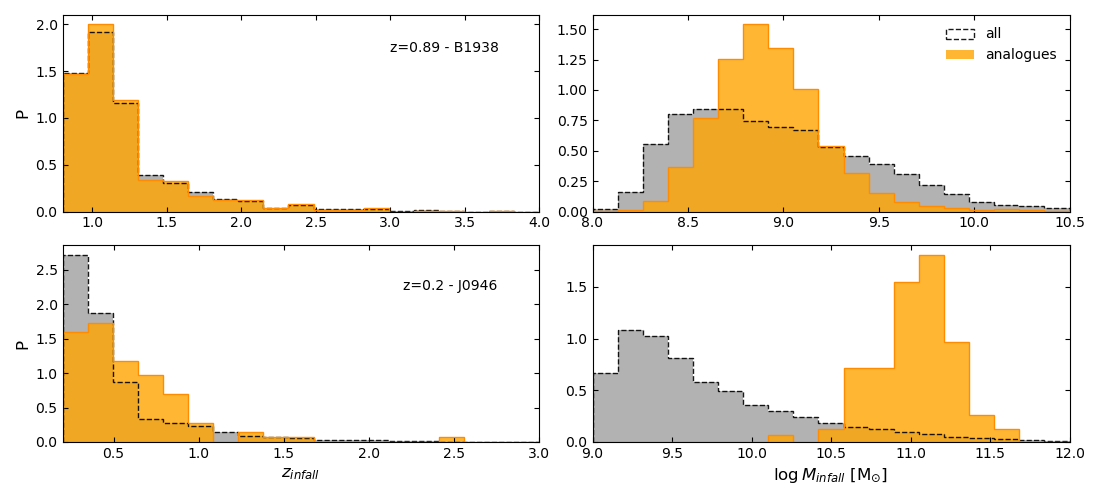}
\caption{Properties of simulated subhaloes at infall. We use the merger trees to track the subhalo evolution and measure their properties at infall, i.e. at the redshift where they first entered their host halo. We plot the normalised distribution of infall redshifts $z_\mathrm{infall}$ (left) and the mass at infall $M_\mathrm{infall}$ (right). The orange histograms show the distribution of the subhaloes selected as analogues, i.e. those lying within the orange boxes in the top panels of Fig. \ref{fig:gamma_2d_vs_m2d_1} for which we show the density profiles in the bottom panels of the same figure. The grey histograms show instead the distribution for all subhaloes in the same host haloes and with final masses between $2\times10^{8}$ M$_{\odot} \leq M_{\rm sub}(z=0.89)\leq 5\times10^{9}$ M$_{\odot}$ for B1938+666 and $10^{9}$ M$_{\odot}\leq M_{\rm sub}(z=0.2)\leq 4\times10^{11}$ M$_{\odot}$ for J0946+1006. }
\label{fig:infall}
\end{figure*}

We then plot the density profiles of the analogues in Fig. \ref{fig:prof} and compare them with the analytical profiles from Fig. \ref{fig:b1938} and \ref{fig:j0946}. As found in many previous works, the population of low-mass subhaloes (left panel, matching the B1938+666 detection) spans a wide range of properties due to the diversity in accretion history and tidal disruption. We also note that, for some systems, the inner part of the density profile may be affected by low particle number, despite the fact that we are well above the spatial resolution limit - we discuss this in more detail in Sect. \ref{subsec:resolution}. In thee right panel, we see that the selected subhalo profiles nicely reproduce the blue curve (i.e. the NFW profile inferred for the J0946+1006 detection) well beyond the region that was used to measure the inner slope. At the same time, the subhaloes present a wide range of truncations in the outskirts, signalling different levels of tidal stripping and, possibly, infall times, as already reported in \citetalias{heinze2023}.
To investigate this hypothesis, we analyse the subhalo merger trees, already described in Sect. \ref{subsec:mnfw}.  For each object, it is possible to walk the tree backwards to determine when it formed as an independent structure and when it then fell into the host halo. Here, we want to compare the infall properties of subhaloes identified as possible analogues of the detections to the general population of subhaloes in the same mass range and at the same redshift. We select the main haloes that host the analogues and follow the merger trees of all their subhaloes in the mass ranges: $2\times10^{8}$ M$_{\odot}\leq M_{\rm sub}(z=0.89)\leq 5\times10^{9}$ M$_{\odot}$ for B1938+666 and $10^{9}$ M$_{\odot}\leq M_{\rm sub}(z=0.2)\leq 4\times10^{11}$ M$_{\odot}$ for J0946+1006. We plot the distribution of infall redshift and mass for the analogues (orange) and the entire population (grey) in Fig. \ref{fig:infall}. For B1938+666, the distribution of infall redshifts z$_{\rm inf}$ is similar for the analogues and the general population, while the infall masses M$_{\rm inf}$ of the analogues shoe a clear peak. The effect is more prominent for the detection in J0946+1006, where we find a difference in the distribution of infall times that excludes some of the most recent mergers, and M$_{\rm inf}$ shows a clear preference towards very high masses. This means that the observed subhaloes are only compatible with descendants of more massive structures that may host a baryonic core and that have been subjected to stripping during their life in the host halo. Both these effects contribute to creating the observed compact mass distributions. However, as seen in Fig. \ref{fig:luminosity}, subhaloes with infall masses corresponding to J0946+1006 are indeed too massive and contain too many stars to be an actual counterpart of the detection.

\begin{figure*}
\includegraphics[width=0.89\columnwidth]{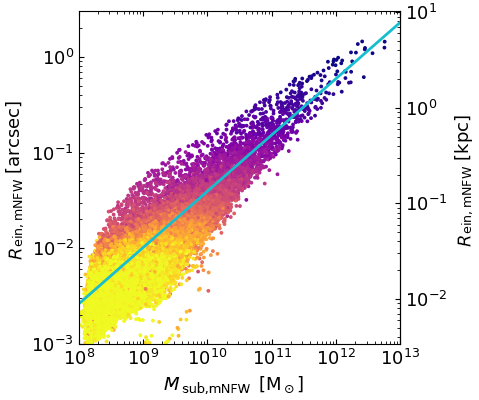}
\hfill
\includegraphics[width=1.11\columnwidth]{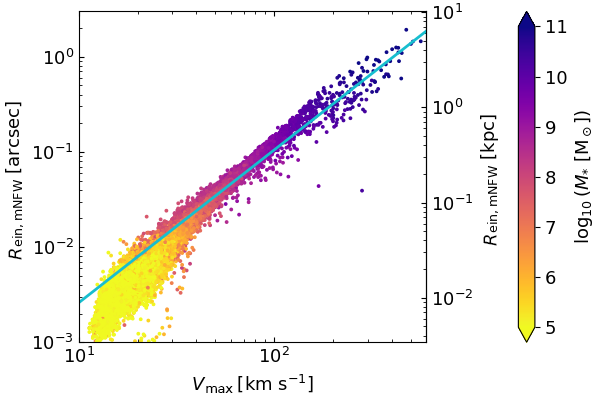}
\caption{The scaling of the Einstein radius (obtained from the best mNFW fits) with subhalo mass (left) and $V_\mathrm{max}$ (right). In both plots, the colour-coding of the data points indicates the subhalo stellar mass $M_{*}$. The cyan lines represent the best power law fit for all of the data points.
}
\label{fig:thetaE_rel_analytic}
\end{figure*}

\begin{figure*}
\includegraphics[width=1.\textwidth]{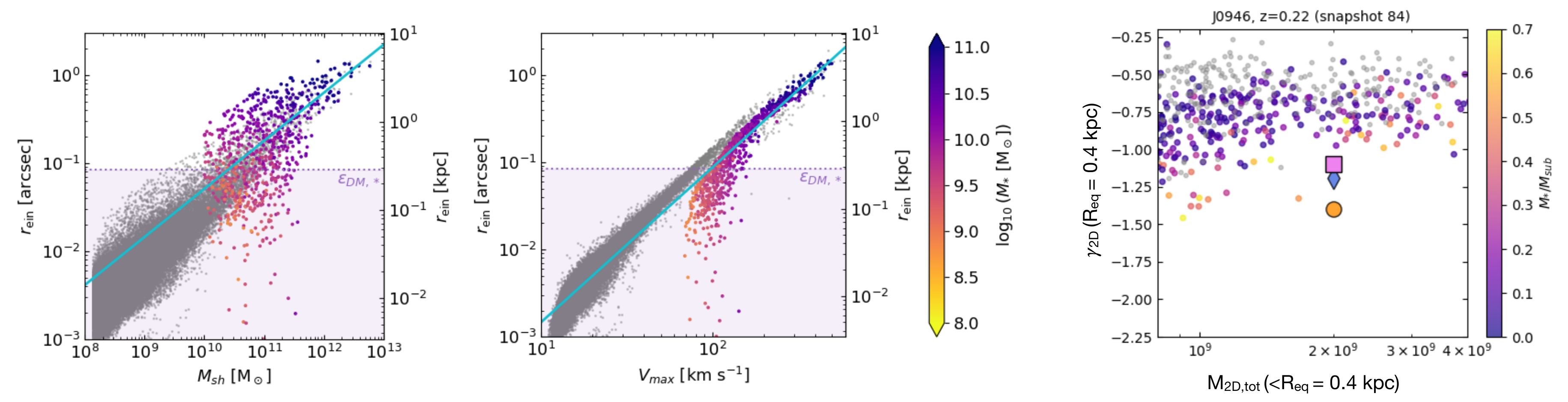}

\caption{Tests of the impact of numerical resolution on our results. \emph{Left and central panel}: comparison between the Einstein radii obtained from the best modified NFW profile fits (in grey - corresponding to the points in Fig. \ref{fig:thetaE_rel_analytic}) and those calculated from the smoothed particle surface density maps, coloured by the subhalo stellar mass. The cyan lines are the best power law fits for the latter. In both plots, the regime below the softening length is coloured purple. $Right$: we repeat the measurements performed for Fig. \ref{fig:gamma_2d_vs_m2d_1}, using $R_{\rm eq}=0.45$ kpc for the lens J0946+1006. 
}
\label{fig:resolution}
\end{figure*}

\subsection{Einstein Radii of the subhaloes} \label{subsec:re}

Given the results of the previous section, it is clear that subhaloes of a given total mass can produce a range of lensing effects, depending on their density profiles. \citetalias{heinze2023} found that the mNFW model describes the average density profile of subhaloes, but individual subhaloes within the same mass bin differ in terms of inner slope and concentration. The spread correlates with other subhalo properties, such as the infall time or the baryonic content. Given that the lensing effect is directly dependent upon the density distribution, it is natural to expect a spread in the lensing properties of subhaloes as well. Here, we extend their results by calculating scaling relations for the subhalo Einstein radius $R_\mathrm{ein}$ as a function of subhalo mass $M_\mathrm{sub}$ and the circular velocity peak $V_\mathrm{max}$. We choose a lensing configuration with the lens (the subhalo) at a redshift of $z_\mathrm{L}=0.2$ and the source at a redshift of $z_\mathrm{s}=0.609$, corresponding to J0946+1006. We do separate analyses for the smoothed particle surface density maps created from $(i)$ the particle data and $(ii)$  the best mNFW profile fits of each subhalo. Here, we present the latter and discuss the potential biases introduced by the spatial and mass resolution of the simulations when using only particle data in Sect. \ref{subsec:resolution}.  To analyse the lensing effect of the analytical mNFW profiles, we integrate the best fitting spherical density profile in the $z$-direction to obtain a surface density profile, and from this, we construct a circular map of $2048^{2}$ pixels with a field of view of $2 r_\mathrm{s}$. In order to resolve the Einstein radii of subhaloes with very different radial extents, the field of view of the ray tracing computations (which is only a part of the total field of view of the entire surface density map) is adaptive - to a minimum of 0.1 arcseconds depending on $r_\mathrm{s}$, and within this field of view $2048^2$ rays are traced. We find the (equivalent) Einstein radius $R_\mathrm{ein}$ by measuring the area enclosed by the tangential critical line and computing the radius of the circle that has the same area, following the approach of \citet{pylenslib}.

Fig. \ref{fig:thetaE_rel_analytic} shows the resulting distribution of $R_\mathrm{ein}$ as a function of subhalo mass and $V_\mathrm{max}$. Here, instead of the subhalo mass $M_\mathrm{sub}$ obtained from the particle data, we compute the subhalo mass $M_\mathrm{sub, mNFW}$ by integrating the analytic profile. In the Appendix, we show that the two mass definitions are equivalent for $M_\mathrm{sub}\leq 10^{11}\mathrm{M_{\odot}}$ (the mass regime we are interested in) and a small bias is introduced at higher masses. With this method, it is now possible to look at subhaloes with low mass and small Einstein radii that would otherwise not be properly resolved due to resolution limitations. 

Despite the clear correlation, we see a significant scatter around the mean at fixed subhalo mass or at fixed $R_\mathrm{ein}$. Gravitational lensing robustly measures the projected mass in the region covered by the lensed images, while the total subhalo mass is inferred via a model for the density profile. These results show how using a mean profile (such as the NFW or the mNFW) may not be sufficient to interpret lensing data, given that a certain Einstein radius could match a wide range of total masses. Moreover, subhaloes with lower $R_\mathrm{ein}$ may simply become non-detectable (although present), and the more lensing-efficient ones will dominate the sample. 
The light blue line represents the best-fit relations. For the top panels, they are given by:
\begin{equation}
    R_\mathrm{ein, mNFW} = (0.0101'' \pm 0.0001'') \left( \frac{M_\mathrm{sub, mNFW}}{10^9 \ \mathrm{M}_\odot} \right)^{0.589 \pm 0.001},
\label{eq:thetaE_m_relation_analytical}
\end{equation}
\vspace{-10pt}
\begin{equation}
    R_\mathrm{ein, mNFW} = (0.105'' \pm 0.001'') \left( \frac{V_{\mathrm{max}}}{100 \ \mathrm{km \ s^{-1}}} \right)^{1.61 \pm 0.01}.
\label{eq:thetaE_vmax_relation_analytical}
\end{equation}

Generally, the scaling relations values also depend on the redshifts of the lens and source. However, the lens and source configuration do not strongly affect the slope of the relations. Because of the $V_\mathrm{max} - M_\mathrm{sub}$ relation, the scaling of $R_\mathrm{ein}$ with $V_\mathrm{max}$ is not surprising, but the scatter of the $R_\mathrm{ein} - V_\mathrm{max}$ relation turns out to be much smaller than for the $R_\mathrm{ein} - M_\mathrm{sub}$ relation.

\subsection{The effect of resolution} \label{subsec:resolution}

It is well known that measurements from simulations can suffer from numerical uncertainties if performed at scales close to the spatial resolution or using low-mass (sub)haloes that are traced by a few particles. Moreover, low-mass subhaloes at the limit of resolution can be subjected to artificial disruption, altering their predicted number and properties \citep{vandenbosch18,green21}. Here, we do not try to correct potential artificial disruption and use the TNG50-1 subhalo catalogues as they are. In this section, we thus present tests that we performed to quantify possible resolution limitations of our analysis (beyond what already discussed in the previous sections) and the strategies to circumvent them. 

As described in Sect. \ref{subsec:re}, we calculated the subhalo Einstein radii both directly from the particle distribution and from the analytical profiles. Using the latter allowed us to circumvent resolution systematics partially and study a larger range of subhalo lensing properties (shown in Fig. \ref{fig:thetaE_rel_analytic}) compared to the results obtained from discrete particle data. For the particle data, we use \textsc{Py-SPHViewer} \citep{pyview} to create surface density maps of $2048^{2}$ pixels for each individual subhalo. For each map, we choose a camera looking at the subhalo from infinity so that the map is rendered using a parallel projection. The field of view is chosen so that every particle of the subhalo is still contained within the map. Based on the position of 32 neighbouring particles, the smoothing length is calculated for each individual particle, and the minimum smoothing length is set to the softening length of TNG50. The density values of each pixel are normalised such that the total mass of the subhalo equals the sum of the pixel values. The side length of the field of view for the ray tracing is set to 2 arcseconds, and $512^2$ rays are traced within it. In Fig. \ref{fig:resolution}, we now compare those measurements (here in grey) to the Einstein radii derived from the particle data. Given the limits imposed by the discreteness of the particle distribution and by the resolution of the simulation, it was, in fact, possible to reliably measure $R_\mathrm{ein}$ only for subhaloes with $M_\mathrm{sub} \geq 10^{10} \ \mathrm{M_\odot}$. In this range, the overall distribution of points looks similar, but the values calculated from the particle distribution have a larger scatter, particularly towards small values of $R_\mathrm{ein}$. In the $R_\mathrm{ein} - V_{\mathrm{max}}$ space, it is particularly clear that the data points below the softening length $\epsilon_\mathrm{DM,*}$ are those that deviate the most from the relation obtained from the analytic profiles. For the analytical profiles, the limitations are instead determined by the number of pixels in the surface density maps, the field of view and the number of rays in the ray tracing computation. These limitations can lead to a small amount of artificial scatter for a few low-mass subhaloes with small Einstein radii, which can be mitigated by increasing the resolution even further. However, this is less significant than the intrinsic resolution limits of the simulations.

A similar resolution effect can also affect our measurements of the projected mass and slope presented in Fig. \ref{fig:gamma_2d_vs_m2d_1}. Here, we repeat the measurement at the redshift of J0946+1006 using $R_{c}=R_{\rm eq}=0.45$ kpc instead of $R_{c}=1$ kpc. As discussed in Sect. \ref{subsec:analogues}, this value is close to the spatial resolution limit of the simulation, and thus our measurement of the projected slope $\gamma_{2D}$, based on five radial bins around $R_{c}$ may be biased. The results are shown in the right panel of Fig. \ref{fig:resolution}, which should be compared to the top-right panel of Fig.  \ref{fig:gamma_2d_vs_m2d_1}. Indeed, the slopes measured at this inner radius are shallower because of the artificial core of the central part of the profiles, highlighting the need to be careful in the treatment of simulations. This happens despite the high stellar mass of the haloes. Considering the difference in the spatial and mass resolution between TNG50-1 and TNG100-1, this measurement is done at a resolution level similar to the results presented in \citet{minor2021}. We thus believe that their reported lack of analogues of the detected subhalo was mostly driven by resolution effects. A similar effect at the low-mass end can be appreciated in the bottom-left panel of Fig. \ref{fig:gamma_2d_vs_m2d_1}, where the resolution limit and the smaller number of particles in subhaloes in the mass range $M_\mathrm{sub}\leq10^{9}$ M$_{\odot}$ conspire to create an even clearer density core. We remark that this happens even though all these objects are defined by a few hundreds of particles.

\section{Discussion and conclusions} \label{sec:conc}

This paper explores the tension between the high concentration of subhaloes detected with gravitational imaging and the predictions from CDM cosmological hydrodynamical simulations. In the first part of the paper, we re-analysed the gravitational lensing observations of two systems, where a perturber has been detected in optical data (HST and Keck-AO): J0946+1006 at $z=0.22$ and B1938+666 at $z=0.881$. We model the lensing signal of the main lenses and the detected subhaloes, exploring four options for the density profile model of the latter: an NFW profile with fixed or variable concentration, a Pseudo-Jaffe (PJ) and a Power-Law (PL)  profile. Here we summarise our main results:

\newcounter{auxCounter}
\begin{enumerate}
      \item We find that all models for the perturbers lead to an increase of Bayesian evidence relative to a smooth model (see Table \ref{tab:models}). Hence, we confirm both detections at a high statistical significance.  We find that three models for the density profile (PJ, PL and NFW with free concentration) are preferred by the data - over the smooth model alone - at about the same significance level. While the NFW, PJ, and PL models agree on the subhalo position within the resolution of the data, the inferred total mass can vary up to $\sim$1 order of magnitude as expected from the profiles. 
    \item The three models that fit the data well predict the same value of projected enclosed mass within a certain characteristic radius $R_{\rm eq}$. This indicates that in these observations, it is not possible to constrain the inner slope of the subhaloes' density profiles - which indeed lie at the limit of the data resolution - while what is important from a lensing point of view is the amount of mass concentrated within $R_{\rm eq}$. This radius of equal enclosed mass could be a more robust estimator for subhalo detections via gravitational lensing, given that it would allow one to bypass the need for specific models of the density and mass profiles.
    \item The detections correspond to extremely dense objects: when interpreted as NFW subhaloes within the main lens, they have concentrations much higher than those predicted by numerical simulations. The best NFW fit is found to have M$_{\rm sub}=(2.026\pm0.401)\times10^{8}$ M$_{\odot}$ and $c_{\rm sub}=256\pm 74$ in B1938+666 and M$_{\rm sub}=(2.23\pm0.23)\times10^{10}$ M$_{\odot}$ and $c_{\rm sub}=201\pm 37$ in J0946+1006. These values are extreme outliers of the concentration-mass-redshift relations measured in dark matter simulations and commonly used to describe NFW profiles (see Fig. \ref{fig:concentration}). Comparing our models, we can thus conclude that the classic NFW profile is not a good description of the detected subhaloes, given that the inner slopes inferred from the lens modelling are close to or steeper than isothermal. This is consistent with previous analyses of the same datasets \citep{minor2021,sengul22,ballard24} and with the results of \citet{heinze2023}, who measured the density profiles of subhaloes in TNG50 and found an average inner slope close to -2.
    \setcounter{auxCounter}{\value{enumi}}

   \end{enumerate}
We then compare our best models to the predictions of the highest-resolution run of the IllustrisTNG sample, i.e. TNG50-1 \citep{pillepich18,nelson19}. \citet{heinze2023} already modelled the density profiles in TNG50-1, finding that the average inner slopes of subhaloes in the hydro run are steeper than the NFW profile and closer to isothermal, albeit with a large scatter. This is due to the combined effect of baryons and tidal stripping and becomes possible because of the high resolution of the simulation, which allows us to measure the density slopes at small scales reliably. We are thus interested in seeing if we can find objects that resemble the two detections among this diversity of profiles. We compare observations and simulations in the parameter space defined by $\gamma_\mathrm{2D}(R_{c})$ and $M_\mathrm{2D}(<R_{c})$, i.e. the average log-slope of the surface density profile around a characteristic radius and the projected mass within the same distance. 

\begin{enumerate}
  \setcounter{enumi}{\value{auxCounter}}
    \item Only $\sim8.8$ per cent of subhaloes in TNG50-1 have inner slopes as steep as the detection in B1938+666, indicating that this is not a \emph{typical} subhalo. However, we are able to find simulated analogues that match the detections in all the considered quantities: slope, enclosed mass, baryonic content and luminosity. These subhaloes are dark-matter dominated, and their density profile spans a wide range of inner slopes that match the values inferred by all three considered profile models. This is consistent with the known spread in the properties of low-mass subhaloes, which are influenced by their tidal history. The $L_{V,\odot}$ luminosities, calculated from the stellar particles of the simulations, are all consistent with the upper limit set by observations. 
    \item The detection in J0946+1006 is more massive, and thus, different profiles produce visibly different lensing properties, such as the critical curves and caustics. In this case, the inferred inner slope is shallower and compatible with $\sim25.6$ per cent of simulated subhaloes. However, the slope is produced by a significant stellar component: in this case, the luminosities of simulated subhaloes are all higher than the upper limit set by the data and thus excluded. Given that this detection is located in a relatively dark part of the lensed arc (see Fig. \ref{fig:lensing_data}), it cannot correspond to a satellite with a large detectable stellar mass. Moreover, the distribution of infall times and infall masses clearly peaks at high values that are consistent with a large satellite. The properties of the detection in J0946+1006 are thus in tension with the properties of subhaloes predicted by CDM hydro simulations - even though here we could find similar slopes and enclosed masses. This indicates either that this system is in tension with the CDM model or that baryonic physics in CDM simulations is currently not able to produce the full range of subhalo properties required by observations.
    \item Finally, we discuss how spatial and mass resolution may influence our results. In J0946+1006, we find more subhaloes that match the slope and enclosed mass of the detection, compared to \citet{minor2021}, thanks to the improved resolution compared to TNG100-1, which allows us to measure the inner slope of the profiles more reliably. We also calculate the Einstein radii of the simulated subhaloes, demonstrating that it is more stable to derive them from the analytical profiles calculated by \citetalias{heinze2023}, while the particle distribution introduces noise. We provide scaling relations between the subhalo mass M$_{\rm sub}$ with $(i)$ the projected enclosed mass of the two detections M$_\mathrm{2D}$ and $(ii)$ the subhalo Einstein radius R$_{\rm ein}$ predicted from the simulations.
\end{enumerate}

In summary, we confirm that the two detections are best explained by extremely compact mass distributions. This is most likely due to the impact of baryonic physics, however, there could be alternative explanations for the measured high concentration based on non-standard dark matter descriptions, such as velocity-dependent self-interacting dark matter (SIDM) models. \citet{nadler21} explored the possibility that a velocity-dependent SIDM model may produce very compact subhaloes even without baryonic effects, finding a better agreement in terms of slope but not of total projected mass. However, our results indicate that the latter is the most important quantity that simulations must reproduce: the results presented in Fig. \ref{fig:b1938} and \ref{fig:j0946} suggest that the considered observations cannot resolve the inner slope of the profile - and thus are not sensitive to it - while they can constrain the projected mass enclosed within $R_{\rm eq}$. 

Here, we have only considered subhaloes within the main lensing galaxy as counterparts of the detections. However, these could also be isolated dark haloes along the line of sight, which are predicted to be more numerous than subhaloes \citep{li16,despali16,amorisco21}. Their properties could differ from those of subhaloes: in particular, we cannot exclude that a halo located further away from the lens could be consistent with the upper limit in luminosity for the J0946+1006 detection. However, it remains to be seen if isolated haloes - that do not suffer from tidal stripping or other interactions that lead to compact mass distributions - can explain the observed properties of the detections. The interpretation of the (sub)halo detections in the context of dark matter physics heavily relies on theoretical models; given that the NFW profile is not a good description for these objects, previous constraints based on classic numerical models may need to be revisited. For instance, if many subhaloes are more concentrated than the classic NFW model, we may ask why we have not detected more dark objects with strong lensing. Deep surveys with the next generation of telescopes will hopefully provide us with new objects over a larger mass range, increasing the statistical sample and revealing more about the dark matter content of the Universe.

\begin{acknowledgements}

GD thanks Devon Powell, Conor O'Riordan, Quinn Minor and Carlo Giocoli for useful discussions and suggestions on lens modelling.
GD acknowledges the funding by the European Union - NextGenerationEU, in the framework of the HPC project – “National Centre for HPC, Big Data and Quantum Computing” (PNRR - M4C2 - I1.4 - CN00000013 – CUP J33C22001170001). CS acknowledges financial support from the Italian National Institute for Astrophysics (INAF, FO: 1.05.12.04.04). S.V. thanks the Max Planck Society for support through a Max Planck Lise Meitner Group. Part of this research was carried out on the High-Performance Computing resources of the FREYA cluster at the Max Planck Computing and Data Facility (MPCDF) in Garching, operated by the Max Planck Society (MPG). RSK acknowledges financial support from the European Research Council via the ERC Synergy Grant ``ECOGAL'' (project ID 855130),  from the Heidelberg Cluster of Excellence (EXC 2181 - 390900948) ``STRUCTURES'', funded by the German Excellence Strategy, and from the German Ministry for Economic Affairs and Climate Action in project ``MAINN'' (funding ID 50OO2206). The team in Heidelberg also thanks {\em The L\"{a}nd} and the German Science Foundation (DFG) for computing resources provided in bwHPC via grants INST 35/1134-1 FUGG and 35/1597-1 FUGG, and also for data storage at SDS@hd funded through grants INST 35/1314-1 FUGG and INST 35/1503-1 FUGG. 

\\\\
The entire data of the IllustrisTNG simulations, including TNG50, are publicly available and accessible at \url{www.tng-project.org/data} \citep{nelson19}. The observational data used here are also publicly available via the HST and Keck archives. This research made use of the public Python packages matplotlib \citep{matplotlib}, NumPy \citep{numpy}, Astropy \citep{astropy:2013,astropy:2018,astropy:2022} and \textsc{COLOSSUS} \citep{diemer18}. 
\end{acknowledgements}

%
   \bibliographystyle{aa} 
   \bibliography{aanda.bbl} 
%

\appendix

\section{Lens models}
\begin{figure*}
\includegraphics[width=\textwidth]{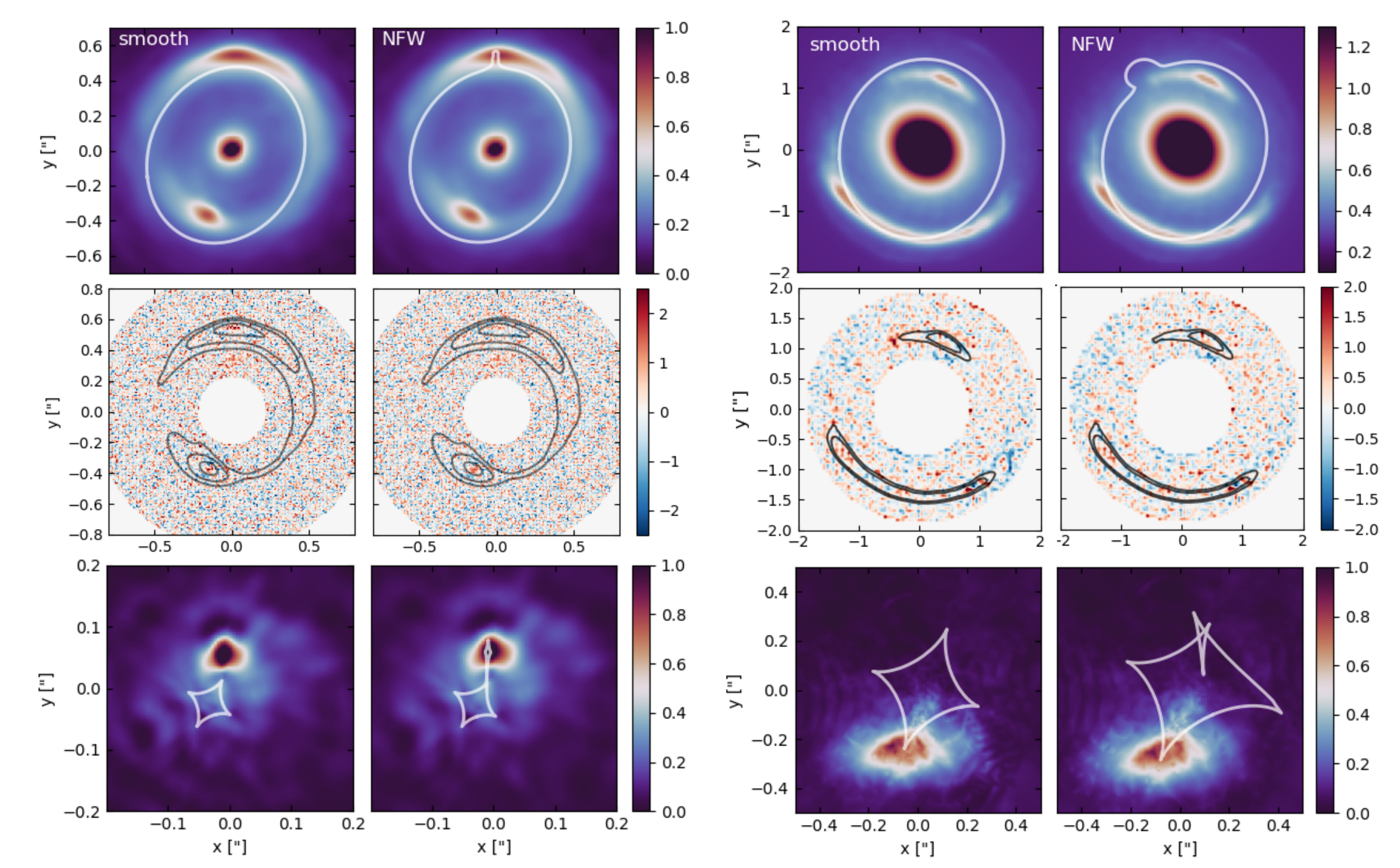}
\caption{Lens modelling results of the two lenses: B1938+666 in the K$^\prime$ Keck band (left) and J0946+1006 in the F814W band (right). For each lens, we present the result of the smooth model and the model that includes the NFW profile with free concentration. In each case, we show $(i)$ the best-reconstructed model and critical lines, including both the light of the central galaxy and the main arc, $(ii)$ the residuals between the data and the best model, calculated as \emph{(data-model)} and $(iii)$ the corresponding best-reconstructed source with the caustics. The posterior distributions for all models are shown in Fig. \ref{fig:lensing_data3} and \ref{fig:lensing_data4}. 
}
\label{fig:lensing_data1}
\end{figure*}

\begin{figure*}
\includegraphics[width=\textwidth]{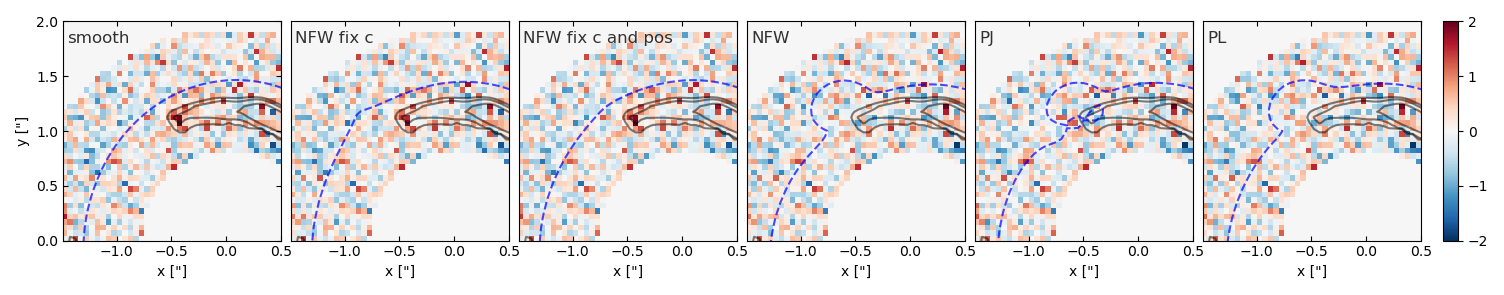}
\caption{We zoom on the detection location in J0946+1006 and show the residuals between the model and the data for all considered variations of subhalo properties. In each panel, the blue dashed line shows the critical curve predicted by the model, and the black contours trace the surface brightness of the arc. In the first panel, we see a clear red spot at the edge of the lensed arc, which is close to the detection location and indicates that the smooth model cannot correctly reproduce the surface brightness distribution. The NFW models with fixed concentration (second panel) and position (third panel) cannot remove this excess, while the residuals reach the noise level when the NFW is allowed to have a higher concentration (fourth panel) or when we use a PJ or PL profile (last two panels).}

\label{fig:lensing_data2}
\end{figure*}

\begin{figure}
\includegraphics[width=\columnwidth]{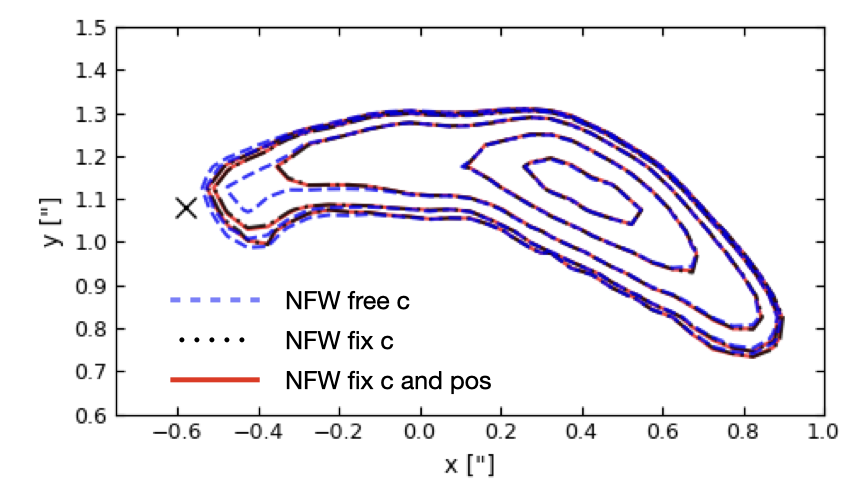}
\caption{Contours showing the surface brightness distribution of the lensed arc of J0946+1006 when modelled with the inclusion of an NFW profile. The three cases shown here correspond to panel 1-4 in Fig. \ref{fig:lensing_data2} and highlight that, among these, only the NFW model with free concentration (blue dashed line) is able to produce the extended shape of the arc's edge which matches the surface brightness of the observes arc.}
\label{fig:lensing_data2b}
\end{figure}

\begin{figure*}
\includegraphics[width=\textwidth]{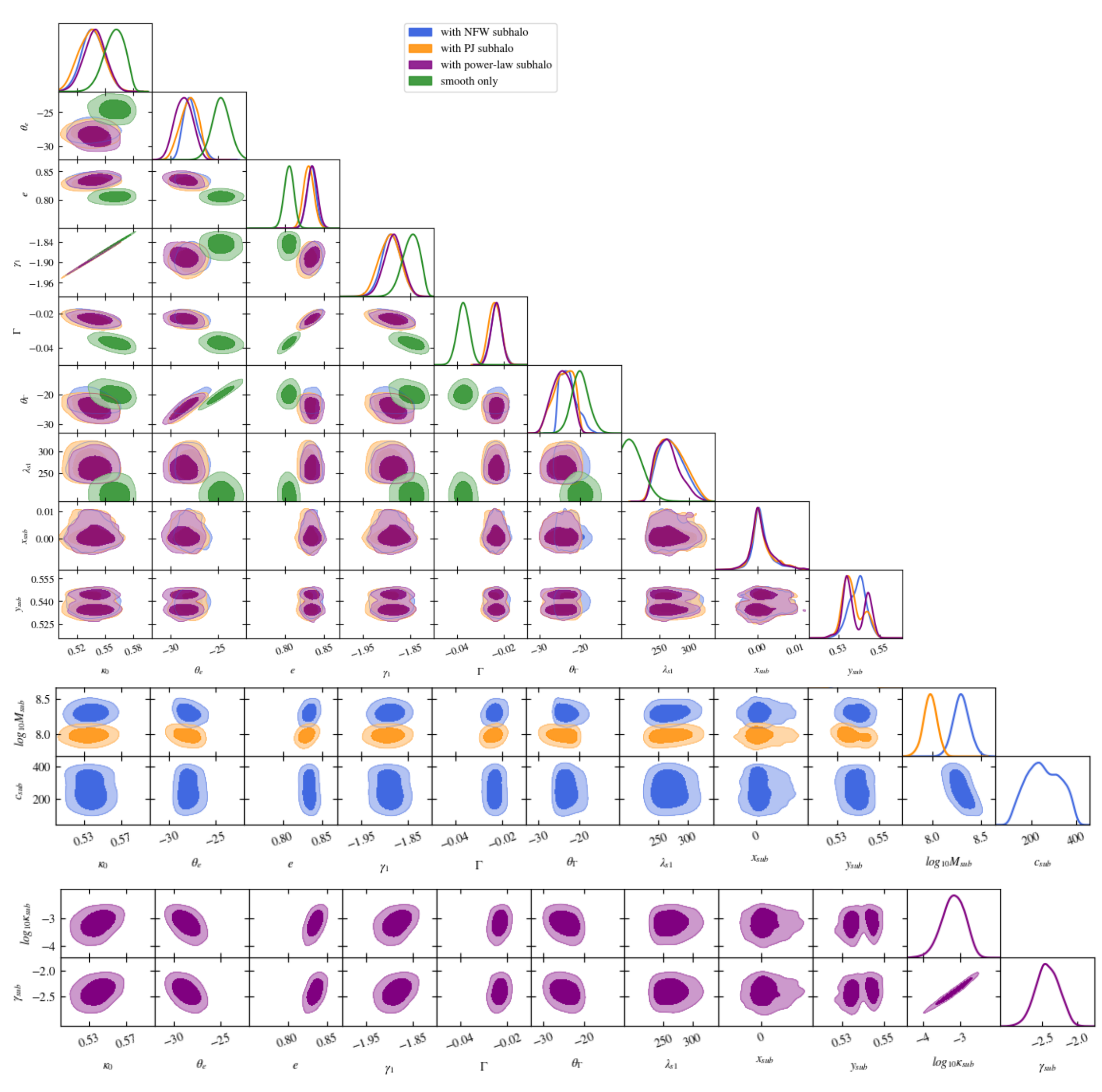}
\caption{Full posterior distributions for the system B1938+666, including the parameters describing the lens, source regularisation and subhalo. We show the smooth model (gree) and the perturbed models that include a subhalo described as PJ, PL or NFW with free concentration.}
\label{fig:lensing_data3}
\end{figure*}

\begin{figure*}
\includegraphics[width=\textwidth]{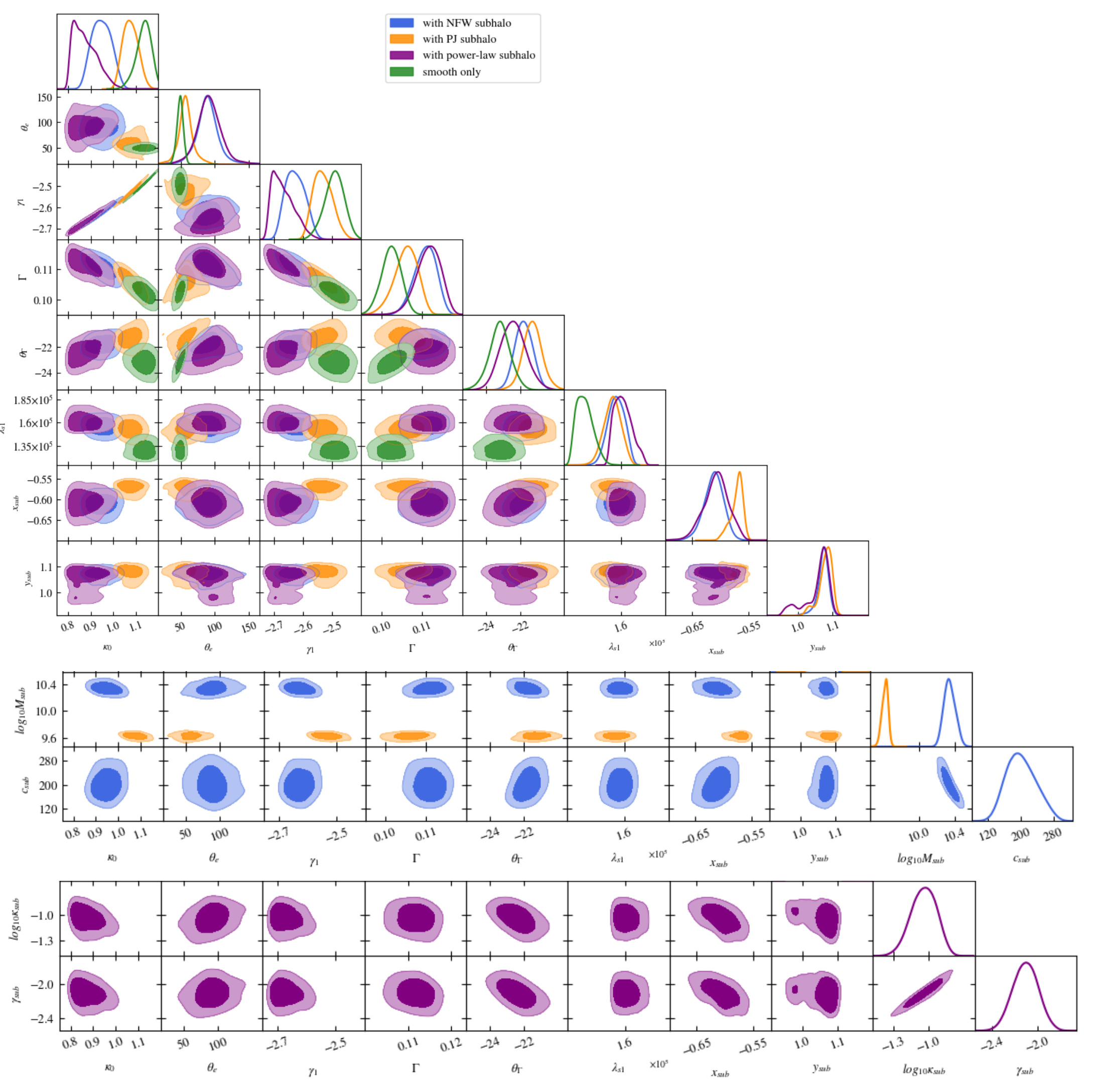}
\caption{Full posterior distributions for the system J0946+1006.}
\label{fig:lensing_data4}
\end{figure*}

For simplicity, in Sect. \ref{sec:results} we show and discuss only the posterior distribution of subhalo parameters calculated with \textsc{Multinest}. In practice, these are optimised simultaneously to infer the parameters describing the main lens and source regularisation. For each model, we thus have the full posterior distribution that can be used to identify degeneracies between the parameters, as well as a mass model and source reconstruction. In Fig. \ref{fig:lensing_data1}, we show some examples of how the inclusion of a perturbed improves the mass model and the residuals. In Fig. \ref{fig:lensing_data3} and Fig. \ref{fig:lensing_data4}, we plot the complete posterior distribution for the smooth model (green) and the three best subhalo models: PJ, PL and NFW with free concentration. The parameters describing the main lens are: the surface mass density normalisation $\kappa_{0}$, the ellipticity $e$, the position angle $\theta_{e}$, the power-law slope $\gamma_1$, the external shear strength $\Gamma$ and its position angle $\theta_{\Gamma}$. The source regularisation is $\lambda_\mathrm{s1}$.  In Figures \ref{fig:lensing_data2} and \ref{fig:lensing_data2b}, we zoom-in on the detection location in J0946+1006 to show which models are able to clearly remove the residuals from the model and that the NFW models with fixed concentration (and position) cannot reproduce the surface brightness of the arc correctly.

\section{Subhalo Mass Bias and additional scaling relations}
\label{sec:mass_bias}
Here we compare the subhalo masses obtained by integrating the best modified NFW profile fits with the masses obtained by summing up the individual masses of all the particles in the simulation that are bound to the subhalo.\\

\begin{figure*}
\centering
\includegraphics[width=\columnwidth]{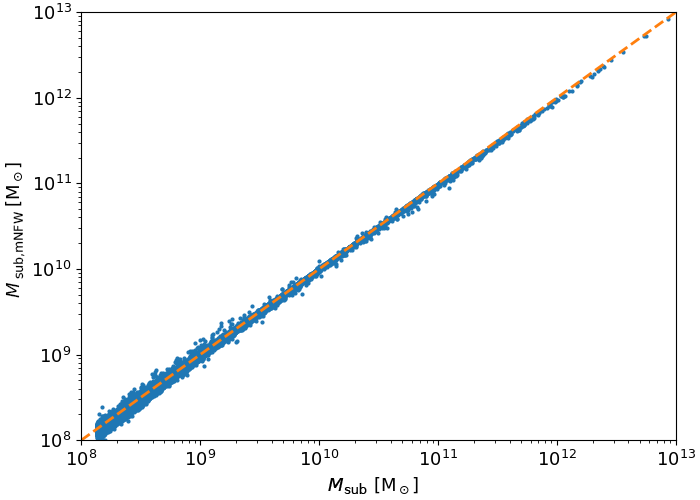}
\includegraphics[width=\columnwidth]{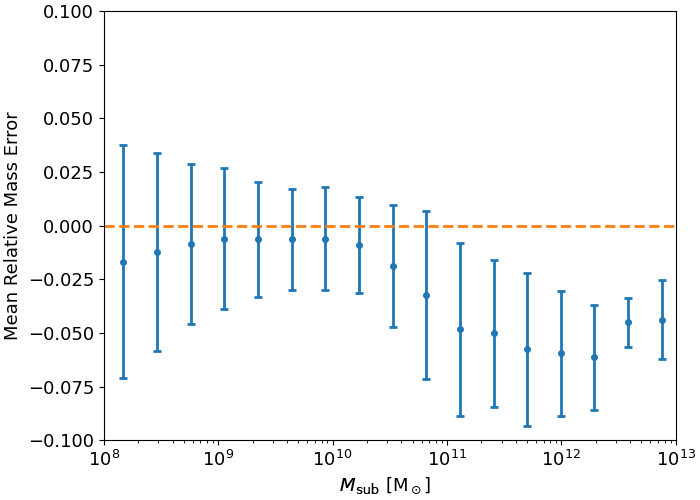}
\includegraphics[width=\columnwidth]{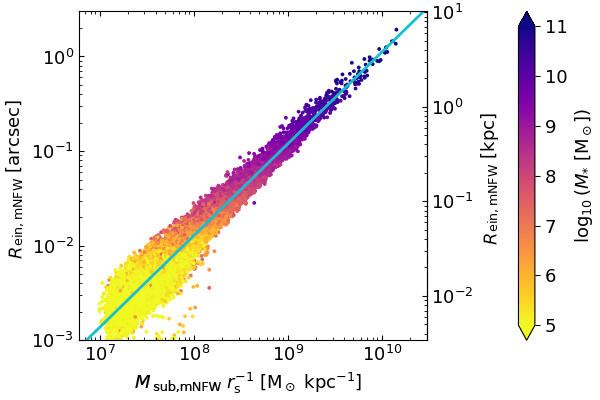}
\caption{
$Top-left$: the subhalo masses obtained by integrating the best modified NFW profile fits are plotted against the subhalo masses obtained from the particle data of the TNG50 simulation. The orange dashed line indicates where both masses are equal. $Top-right$: the mean relative errors of the subhalo masses obtained by integrating the best modified NFW profile fits are plotted against the true subhalo masses obtained from the simulation data for each mass bin. The standard deviations are indicated by the error bars. $Bottom$: Scaling of the Einstein radius (obtained by using the best modified NFW profile fits) with $M_\mathrm{sub, mNFW} \ r_\mathrm{s}^{-1}$. The colour-coding indicates the concentration $c_V$. The blue line shows the best power law fit.
}
\label{fig:subhalo_mass_bias}
\end{figure*}

The top panels of Fig. \ref{fig:subhalo_mass_bias} illustrate the results of this analysis. In the left panel, the subhalo masses obtained by integrating the best modified NFW profile are plotted against those obtained from the particle data for all the subhaloes in the TNG50 sample. Here, one can already see that the modified NFW profile can accurately reproduce the actual subhalo masses with only small errors. 
Overall, the modified NFW profile is biased in the sense that the inferred masses are, on average, underestimated. This is especially true for subhaloes with $M_{\rm sub} \geq 10^{11}$ M$_\odot$, where the most likely cause for the underestimation is the fact that the central bumps with an increased density are not parametrised by the model. 

For 79.4 per cent of the subhaloes, the deviation from the true mass is less than 5 per cent, and for 26.3 per cent of the subhaloes, the deviation is less than 1 per cent. The right panel shows in more detail how much the masses are underestimated in each mass bin. One can see that the best mass estimates are obtained for subhaloes with masses in the range $10^9$ M$_\odot$ $ \leq M_{\rm sub} \leq 10^{10}$ M$_\odot$, which is close to the range in which the modified NFW profile provides the best fits for the individual and average subhalo density profiles.

In Sect. \ref{subsec:re}, we calculated linear scaling relations between the Einstein radius $R_{\mathrm ein}$, the subhalo mass and circular velocity. Another relation with a small scatter can be found when looking at how the Einstein radius obtained from the best mNFW profile fit $R_\mathrm{ein, mNFW}$ scales with $M_\mathrm{sub, mNFW} \ r_\mathrm{s}^{-1}$ (see the bottom panel of Fig. \ref{fig:subhalo_mass_bias}). The best power law fit for this relation is given by:
\begin{equation}
    R_\mathrm{ein, mNFW} = (0.0147'' \pm 0.0001'') \left( \frac{M_\mathrm{sub, mNFW} \ r_\mathrm{s}^{-1}}{10^8 \ \mathrm{M}_\odot \ \mathrm{kpc}^{-1}} \right)^{0.97 \pm 0.01}.
\label{eq:thetaE_m_rs_relation_analytical}
\end{equation}

The existence of this scaling relation can be easily explained using a scaling relation for the $r_\mathrm{s}$ parameter of the modified NFW profile from \cite{heinze2023}:

\begin{equation}
    r_\mathrm{s} = (10.2 \pm 0.1) \ \mathrm{kpc} \ \left( \frac{M_{\rm sub}}{10^{10} \ \mathrm{M_\odot}} \right)^{0.9 \pm 0.1} \left( \frac{V_{\mathrm{max}}}{10^2 \ \mathrm{km \ s^{-1}}} \right)^{-1.6 \pm 0.1}.
\label{eq:rs_vmax_fit}
\end{equation}
According to this equation, dividing the subhalo mass by $r_\mathrm{s}$ leads to the approximate cancellation of $M_\mathrm{sub}$ and $V_{\mathrm{max}}^{1.6}$ is left. This results in the scaling we find in the $R_\mathrm{ein, mNFW} - V_{\mathrm{max}}$ relation in equation (\ref{eq:thetaE_vmax_relation_analytical}) and therefore the exponent in equation (\ref{eq:thetaE_m_rs_relation_analytical}) is approximately equal to 1.

\end{document}